\documentclass{bmcart}

\usepackage[utf8]{inputenc} %unicode support
\usepackage{graphicx} %NF added
\usepackage{lineno} %NF added
\usepackage{cite} %NF added

\startlocaldefs
\endlocaldefs

\begin{document}
	
	%%% Start of article front matter
	\begin{frontmatter}
		
		\begin{fmbox}
			\dochead{Research}
			
			
			\title{Analytical estimation of maximum fraction of infected individuals with one-shot non-pharmaceutical intervention in a hybrid epidemic model}
			
			%%%%%%%%%%%%%%%%%%%%%%%%%%%%%%%%%%%%%%%%%%%%%%
			%%                                          %%
			%% Enter the authors here                   %%
			%%                                          %%
			%% Specify information, if available,       %%
			%% in the form:                             %%
			%%   <key>={<id1>,<id2>}                    %%
			%%   <key>=                                 %%
			%% Comment or delete the keys which are     %%
			%% not used. Repeat \author command as much %%
			%% as required.                             %%
			%%                                          %%
			%%%%%%%%%%%%%%%%%%%%%%%%%%%%%%%%%%%%%%%%%%%%%%
			
			\author[
			addressref={aff1,aff2,aff3},                   % id's of addresses, e.g. {aff1,aff2}
			corref={aff1},                       % id of corresponding address, if any
			%noteref={n1},                        % id's of article notes, if any
			email={fujiwara@se.is.tohoku.ac.jp}   % email address
			]{\inits{NF}\fnm{Naoya} \snm{Fujiwara}}
			\author[
			addressref={aff4,aff1}
			%   email={john.RS.Smith@cambridge.co.uk}
			]{\inits{TO}\fnm{Tomokatsu} \snm{Onaga}}
			\author[
			addressref={aff5}
			%   email={john.RS.Smith@cambridge.co.uk}
			]{\inits{TW}\fnm{Takayuki} \snm{Wada}}
			\author[
			addressref={aff6}
			%   email={john.RS.Smith@cambridge.co.uk}
			]{\inits{ST}\fnm{Shouhei} \snm{Takeuchi}}
			\author[
			addressref={aff7}
			%   email={john.RS.Smith@cambridge.co.uk}
			]{\inits{JS}\fnm{Junji} \snm{Seto}}
			\author[
			addressref={aff8}
			%   email={john.RS.Smith@cambridge.co.uk}
			]{\inits{TN}\fnm{Tomoki} \snm{Nakaya}}
			\author[
			addressref={aff9},
			email={john.RS.Smith@cambridge.co.uk}
			]{\inits{KA}\fnm{Kazuyuki} \snm{Aihara}}
			%%%%%%%%%%%%%%%%%%%%%%%%%%%%%%%%%%%%%%%%%%%%%%
			%%                                          %%
			%% Enter the authors' addresses here        %%
			%%                                          %%
			%% Repeat \address commands as much as      %%
			%% required.                                %%
			%%                                          %%
			%%%%%%%%%%%%%%%%%%%%%%%%%%%%%%%%%%%%%%%%%%%%%%
			
			\address[id=aff1]{%                           % unique id
				\orgname{Graduate School of Information Sciences, Tohoku University}, % university, etc
				\street{6-3-09 Aoba, Aramaki-aza Aoba-ku},                     %
				\postcode{980-8579}                                % post or zip code
				\city{Sendai, Miyagi},                              % city
				\cny{Japan}                                    % country
			}
			\address[id=aff2]{%
				\orgname{Institute of Industrial Science, The University of Tokyo},
				\street{4-6-1 Komaba},
				\postcode{153-8505}
				\city{Meguro-ku, Tokyo},
				\cny{Japan}
			}
			\address[id=aff3]{%
				\orgname{Center for Spatial Information Science, The University of Tokyo},
				\street{5-1-5 Kashiwanoha},
				\postcode{277-8508}
				\city{Kashiwa, Chiba},
				\cny{Japan}
			}
			\address[id=aff4]{%
				\orgname{Frontier Research Institute for Interdisciplinary Sciences, Tohoku University},
				\street{Aramaki aza Aoba 6-3, Aoba-ku},
				\postcode{980-8578}
				\city{Sendai, Miyagi},
				\cny{Japan}
			}
			\address[id=aff5]{%
				\orgname{Graduate School of Human Life Science Human Life Science Course, Osaka City University},
				\street{3-3-138, Sugimoto, Sumiyoshi-ku},
				\postcode{558-8585}
				\city{Osaka, Osaka},
				\cny{Japan}
			}
			\address[id=aff6]{%
				\orgname{Faculty of Nursing and Nutrition, University of Nagasaki},
				\street{1-1-1 Manabino, Nagayo-cho, Nishi-Sonogi-gun},
				\postcode{851-2195}
				\city{Nagasaki},
				\cny{Japan}
			}
			\address[id=aff7]{%
				\orgname{Department of Microbiology, Yamagata Prefectural Institute of Public Health},
				\street{1-6-6 Toka-machi},
				\postcode{990-0031}
				\city{Yamagata, Yamagata},
				\cny{Japan}
			}
			\address[id=aff8]{%
				\orgname{Graduate School of Environmental Studies, Tohoku University},
				\street{Aoba, 468-1, Aramaki, Aoba-ku},
				\postcode{980-8572}
				\city{Sendai, Miyagi},
				\cny{Japan}
			}
			\address[id=aff9]{%
				\orgname{International Research Center for Neurointelligence, The University of Tokyo},
				\street{7-3-1 Hongo},
				\postcode{113-0033}
				\city{Bunkyo-ku, Tokyo},
				\cny{Japan}
			}
			%%%%%%%%%%%%%%%%%%%%%%%%%%%%%%%%%%%%%%%%%%%%%%
			%%                                          %%
			%% Enter short notes here                   %%
			%%                                          %%
			%% Short notes will be after addresses      %%
			%% on first page.                           %%
			%%                                          %%
			%%%%%%%%%%%%%%%%%%%%%%%%%%%%%%%%%%%%%%%%%%%%%%
			
			\begin{artnotes}
				%\note{Sample of title note}     % note to the article
				%\note[id=n1]{Equal contributor} % note, connected to author
			\end{artnotes}
			
		\end{fmbox}% comment this for two column layout
		
		%%%%%%%%%%%%%%%%%%%%%%%%%%%%%%%%%%%%%%%%%%%%%%
		%%                                          %%
		%% The Abstract begins here                 %%
		%%                                          %%
		%% Please refer to the Instructions for     %%
		%% authors on http://www.biomedcentral.com  %%
		%% and include the section headings         %%
		%% accordingly for your article type.       %%
		%%                                          %%
		%%%%%%%%%%%%%%%%%%%%%%%%%%%%%%%%%%%%%%%%%%%%%%
		
		\begin{abstractbox}
			
			\begin{abstract} % abstract
				\parttitle{Background} %if any
				Facing a global epidemic of new infectious diseases such as COVID-19, non-pharmaceutical interventions (NPIs), which reduce transmission rates without medical actions, are being implemented  around the world to mitigate spreads. One of the problems in assessing the effects of NPIs is that different NPIs have been implemented at  different times based on the situation  of each country; therefore, few assumptions can be shared about how the introduction of policies affects the patient population. Mathematical models can contribute to further understanding these phenomena by obtaining analytical solutions as well as numerical simulations.
				
				\parttitle{Methods and Results} %if any
				In this study, an NPI was introduced into the SIR model for a conceptual study of infectious diseases under the condition that the transmission rate was reduced to a fixed value only once within a finite time duration, and its effect was analyzed numerically and theoretically. It was analytically shown that the maximum fraction of infected individuals and the final size could be larger if the intervention starts too early. The analytical results also suggested that more individuals may be infected at the peak of the second wave with a stronger intervention.
				
				\parttitle{Conclusions}
				This study provides quantitative relationship between the strength of a one-shot intervention and the reduction in  the number of patients with no approximation. This suggests  the importance of the strength and time of NPIs, although detailed studies are necessary for the implementation of  NPIs in complicated real-world environments as the model used in this study is based on various simplifications.
				
			\end{abstract}
			
			%%%%%%%%%%%%%%%%%%%%%%%%%%%%%%%%%%%%%%%%%%%%%%
			%%                                          %%
			%% The keywords begin here                  %%
			%%                                          %%
			%% Put each keyword in separate \kwd{}.     %%
			%%                                          %%
			%%%%%%%%%%%%%%%%%%%%%%%%%%%%%%%%%%%%%%%%%%%%%%
			
			\begin{keyword}
				\kwd{infectious diseases}
				\kwd{pandemics}
				\kwd{non-pharmaceutical interventions}
				\kwd{hybrid dynamical systems}
			\end{keyword}
			
			% MSC classifications codes, if any
			%\begin{keyword}[class=AMS]
			%\kwd[Primary ]{}
			%\kwd{}
			%\kwd[; secondary ]{}
			%\end{keyword}
			
		\end{abstractbox}
		%
		%\end{fmbox}% uncomment this for twcolumn layout
		
	\end{frontmatter}
	
	%\linenumbers
	
	%%%%%%%%%%%%%%%%%%%%%%%%%%%%%%%%%%%%%%%%%%%%%%
	%%                                          %%
	%% The Main Body begins here                %%
	%%                                          %%
	%% Please refer to the instructions for     %%
	%% authors on:                              %%
	%% http://www.biomedcentral.com/info/authors%%
	%% and include the section headings         %%
	%% accordingly for your article type.       %%
	%%                                          %%
	%% See the Results and Discussion section   %%
	%% for details on how to create sub-sections%%
	%%                                          %%
	%% use \cite{...} to cite references        %%
	%%  \cite{koon} and                         %%
	%%  \cite{oreg,khar,zvai,xjon,schn,pond}    %%
	%%  \nocite{smith,marg,hunn,advi,koha,mouse}%%
	%%                                          %%
	%%%%%%%%%%%%%%%%%%%%%%%%%%%%%%%%%%%%%%%%%%%%%%
	
	%%%%%%%%%%%%%%%%%%%%%%%%% start of article main body
	% <put your article body there>
	
	%%%%%%%%%%%%%%%%
	%% Background %%
	%%
	%\section*{Content}
	%Text and results for this section, as per the individual journal's instructions for authors. %\cite{koon,oreg,khar,zvai,xjon,schn,pond,smith,marg,hunn,advi,koha,mouse}
	
	\section*{Introduction}
	Because of the global spread of COVID-19, our human society is facing a major public health crisis. The COVID-19 pandemic is caused by an emerging pathogen, SARS-CoV-2, for which there is no immunized population, causing an overshooting increase in the number of infected patients and depleting medical resources in many countries. Medical institutions are facing a difficult situation in which they must control second transmissions while treating critically ill patients, and as the number of patients increases, the medical system becomes swiftly tighter. When the number of patients exceeds the capacity, the quality of medical care deteriorates drastically, and the number of medical devices required for life support reaches its limit. This situation further increases the fatality rate of this infectious disease and causes serious damage to our society.
	
	No effective treatment for COVID-19 has been established yet as of September, 2020, and only a public health approach can function as a control measure for the epidemic. To mitigate the spread of COVID-19, each country is implementing non-pharmaceutical interventions (NPIs) \cite{alw07} to regulate social activities. NPIs comprise policies such as case isolation, voluntary home quarantine, closure of schools and universities, social distancing, stopping mass gatherings, and border closure. In major European countries, these NPIs were implemented, depending on the epidemic situation, during the first part of spring in 2020,  with different timings and intensities \cite{ck20,fmg20-13}. Although the first phase of the epidemic appeared to be suppressed by these mitigation measures, the re-epidemic became clearer in almost all countries because of deregulation after the first epidemic.
	
	Because the control of epidemics by NPIs has caused a situation involving the imposition of strong restrictions on human socioeconomic activities, it would be desirable to study in advance the optimal timing, intensity, and duration of interventions that could bring about more promising results with minimal damage to society. Regarding COVID-19, for the purpose of ex-post verification, the influence of NPIs implemented in each country on the effective reproduction number was estimated. In European countries, the correlation between the decay of the effective reproduction number and the implementation of various NPIs has been verified \cite{fmg20-13}. The impact of travel limitations in China on the spread of infection has been discussed \cite{cdg20,kyg20}. However, a reliable estimation of the effects of NPIs is difficult, because the differences in NPI strategies employed by each country are strongly related to various background factors, such as the epidemic situations, social structure, legal systems, and culture \cite{ck20}.
	
	Mathematical modelling is an important method for estimating the effect of NPIs. In particular, the global pandemic of COVID-19 revealed that a situation in which only NPIs are effective against an emerging infectious disease is possible in societies in the 2020s. Such a situation
	% in which only NPIs were effective for emerging infectious diseases 
	had been ``neglected" as a practical research target, which enhances the importance of theoretical approaches. Recently, compartmental models such as the SIR model \cite{km27,am92,cb09,i17} have been extended to estimate the effects of NPIs on the number of patients \cite{fsj10,map11,hdd13,wrk05}. In other settings, optimal policies where intervention intensity can change continuously over time have been discussed in the context of minimizing  objective functions \cite{lml10,pcc14,zyz14,k20}. These solutions are very useful if we can estimate the effect of NPIs on the change in transmission parameters precisely.
	
	The impacts of NPIs should be evaluated as a discontinuous change of the transmission rate in a model to represent the temporal discontinuity of intervention in the real world. The dynamics of the number of infected patients in continuous time in a system that includes discrete parameters, state spaces, and continuous-time dynamics can be modeled as a {\it hybrid dynamical system} \cite{f88,ll09,as10,dh10,hdl10}. This framework has been applied to  mathematical models of infectious diseases \cite{acm93,wxc14,ckr20}. In the simplest case, the {\it one-shot intervention} model, in which the intervention is implemented only once during the epidemic, can be used to discuss the theoretical dependence of the effects of NPIs on the timing and intensity \cite{bf07}. As the COVID-19 epidemic continues, the accumulation of theoretical research on the effects brought about by NPIs has become even more significant. Recently, compartmental models with intervention have been studied both numerically \cite{ahh20,dkm20} and using some analytical methods \cite{mrp20,sgs20} in line with the COVID-19 epidemic.
	
	In this study, we provide exact solutions of a simple SIR model with one-shot intervention, represented by a single discrete reduction in the transmission parameter during an epidemic. These solutions describe the dependence of the peak number of infected patients on the reproduction number under consideration of the implementation of NPIs and intervention timing.  Theoretical and numerical analyses revealed non-trivial relations among the intensity of suppression of pandemics via NPIs, the number of infected individuals at the peaks, and the final size of infection cases.
	%, which may provide a suggestive knowledge about mitigation of the infection by NPIs. 
	The methods and results shown in this study provide basic theoretical understanding in the context of the evaluation of NPIs.

	%%%%%%%%%%%%%%%%%%%%%%%%%%%%%%%%%%%%
	\section*{Materials and methods}
	
	In this study, we focus on the dependence of the maximum fraction of infected individuals on the timing of the NPIs. Note that this analysis is motivated by the COVID-19 epidemic, but we consider a hypothetical epidemic, whose properties are not necessarily the same as that of COVID-19.

	\subsection*{Model}
	We here introduce the SIR model, where the time evolution of the fraction of susceptible ($s(t)$), infected ($i(t)$), and  removed ($r(t)$) individuals is given by the following ordinary differential equation (ODE):
	\begin{eqnarray}
		\frac{ds}{dt} &=& -\beta s i, \label{eq:s}\\
		\frac{di}{dt} &=& \beta s i - \gamma i, \label{eq:i} \\
		\frac{dr}{dt} &=& \gamma i, \label{eq:r}
	\end{eqnarray}
	%in order to discuss the maximum of the fraction of the infected individuals and the final size of the outbreak
	taking into account the one-shot intervention and the second wave after the intervention.
	For simplicity, the total population is assumed to be unity, that is, $s(t)+i(t)+r(t)=1$ holds. The summation of the right-hand sides of Eqs.~(\ref{eq:s})--(\ref{eq:r}) vanishes, which guarantees conservation of the total population. Therefore, as seen below, only Eqs.~(\ref{eq:s}) and (\ref{eq:i}) are numerically integrated to obtain the time evolutions of $s(t)$, $i(t)$, and $r(t)$. 
	% \begin{align}
	%     s(t)+i(t)+r(t)=1. \label{eq:conserve}
	% \end{align}
	In this model, the basic reproduction number and the effective reproduction number at time $t$ are given by $R_0 = \beta/\gamma $ and $R_t = \beta s(t) / \gamma$, respectively.

	Kermack and McKendrick \cite{km27} derived the equation for the maximum fraction of infected individuals and showed that  the final size is obtained by solving a transcendental equation with a given $R_0$ in the SIR model without any intervention \cite{am92,cb09,i17}. By applying the technique they employed in the derivation, we here provide the relationship between the fractions of susceptible and removed individuals at arbitrary times $t_0$ and $t_1$, that is, $s(t_0)$, $r(t_0)$, and $s(t_1)$, $r(t_1)$ as
	\begin{eqnarray}
		s(t_1) &=& s(t_0) \exp\bigg\{ - \frac{\beta}{\gamma} [r(t_1)-r(t_0)] \bigg\}. \label{eq:s0_t}
	\end{eqnarray}
	See Supplementary Section 1, Additional File 1 for details of the derivation.
	All analytical results presented in this paper are derived based on this equality.

	\subsection*{Non-pharmaceutical intervention}
	In the present framework, an NPI in an isolated population is represented by a change in the transmission rate $\beta$. Let $\beta = \beta_{\rm off} (>\gamma)$ be the transmission rate without the intervention, and it is switched to $\beta_{\rm on} (<\beta_{\rm off})$  when the intervention starts at $t=t_{\rm on}$, and restored to $\beta_{\rm off}$ at $t=t_{\rm off}=t_{\rm on}+\Delta t$.
	Let the corresponding basic reproduction numbers be $R_{0,\rm off}=\beta_{\rm off}/\gamma$ and $R_{0,\rm on}=\beta_{\rm on}/\gamma$, respectively.
	%Note that Eqs.~(\ref{eq:s})-(\ref{eq:r}) are invariant under the transformations $\beta \rightarrow k \beta$, $\gamma \rightarrow k \gamma$, $t \rightarrow t/k$.
	Here, $\Delta t$ denotes the duration of the intervention.
	% In this paper, we study the effect of {\it one-shot intervention}, the case where the intervention is implemented only once.
	In this paper, we study the effect of one-shot intervention, one of the simplest implementation schemes, where the intervention is implemented only once. The mathematical methods employed in this study can be applied to more complex cases, for example, where multiple interventions are implemented intermittently.
	If the fraction of susceptible individuals remains large enough and the herd immunity is not achieved after the intervention, a second wave occurs. It is thus necessary to consider this second wave to evaluate the effect of the intervention.
	
	The one-shot intervention setting has been numerically studied by Bootsma and Ferguson \cite{bf07}, showing that the final size, which depends on the timing $t_{\rm on}$ and $t_{\rm off}$, can be smaller for weaker intervention with larger $R_{0,\rm on}$. Di Lauro et al. \cite{dkm20} numerically studied the dependence of the final size and the peak fraction of infected individuals on the timing of the intervention, where the intervention was assumed to start when the fraction of infected and recovered individuals exceeds a threshold value. In particular, they concluded that the onset timing should be chosen so that the two peaks during and after the intervention are comparable.
	Morris et al. \cite{mrp20} showed that the maximum fraction of infected individuals with one-shot interventions can approach that achieved by the optimal intervention, which requires an unrealistic intervention, such as $R_{0,\rm on}=0$.
	Sadeghi et al. \cite{sgs20} also suggested the existence of the optimal timing of the intervention based on discussion using a linearized equation and numerical simulation. 
	%This is an example of the hybrid dynamical systems, which consist of continuous and discrete time evolution \textcolor{red}{[REF]}. 
	
	\section*{Analysis}
\begin{figure}[ht!]
		\centering
		\includegraphics[width=0.95\textwidth]{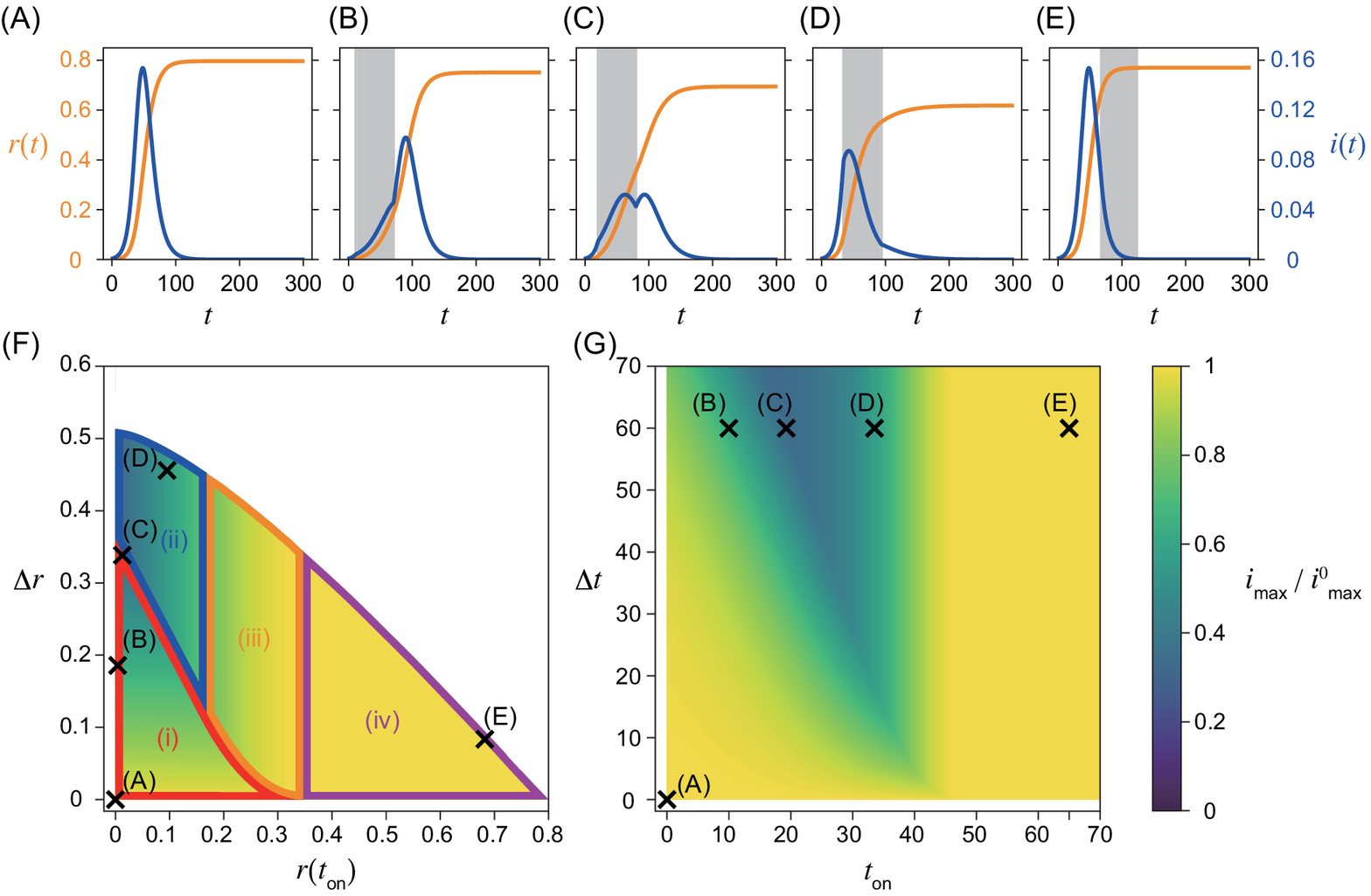}
		\caption{\csentence{Weak intervention with large $R_{0,\rm on}=1.4>1$.} Parameters are $\beta_{\rm off}=2/7 \ {\rm days}^{-1}$ and $\gamma=1/7 \ {\rm days}^{-1}$. 
			(A) Time series of the fraction of infected and removed individuals, $i(t)$ in blue and $r(t)$ in orange, without the intervention. Time series with the intervention with $\beta_{\rm on}=1.4/7 \ {\rm days}^{-1}$ and the intervention duration, $\Delta t=60$ days, with onset times (B) $t_{\rm on}= 10$ days, (C) $t_{\rm on}= 19.2$ days, (D) $t_{\rm on}=33.4$ days, and (E) $t_{\rm on}=65$ days, are depicted. The intervention is implemented for $t_{\rm on}\le t \le t_{\rm off}=t_{\rm on}+\Delta t$ and are represented by grey intervals in these panels. Conditions for the peaks of the fraction of infected individuals are given by $s(t)=1/R_{0,\rm off}$, when the intervention is not implemented, and by $s(t)=1/R_{0,\rm on}$ during intervention.
			The final size of the outbreak for each case is represented by $r(\infty)$. Panels (F) and (G) represent the maximum fraction of infected individuals $i_{\rm max}$ normalized by that without the intervention $i_{\rm max}^0$, plotted in terms of $r(t_{\rm on})$ and $\Delta r = r(t_{\rm off})-r(t_{\rm on})$, and $t_{\rm on}$ and $\Delta t$, respectively. Symbols (A)-(E) in panels (F) and (G) denote the intervention timings of the time series of the corresponding panels in (A)-(E). Symbols (i)-(iv) in panel (F) denote the timing that the maximum infected fraction is observed, as described in the main text. The boundaries between regions shown in panel (F) are obtained analytically. See the main text for the details. 
		}
		\label{fig:large_beta1}
	\end{figure}

\begin{figure}
		\centering
		\includegraphics[width=0.95\textwidth]{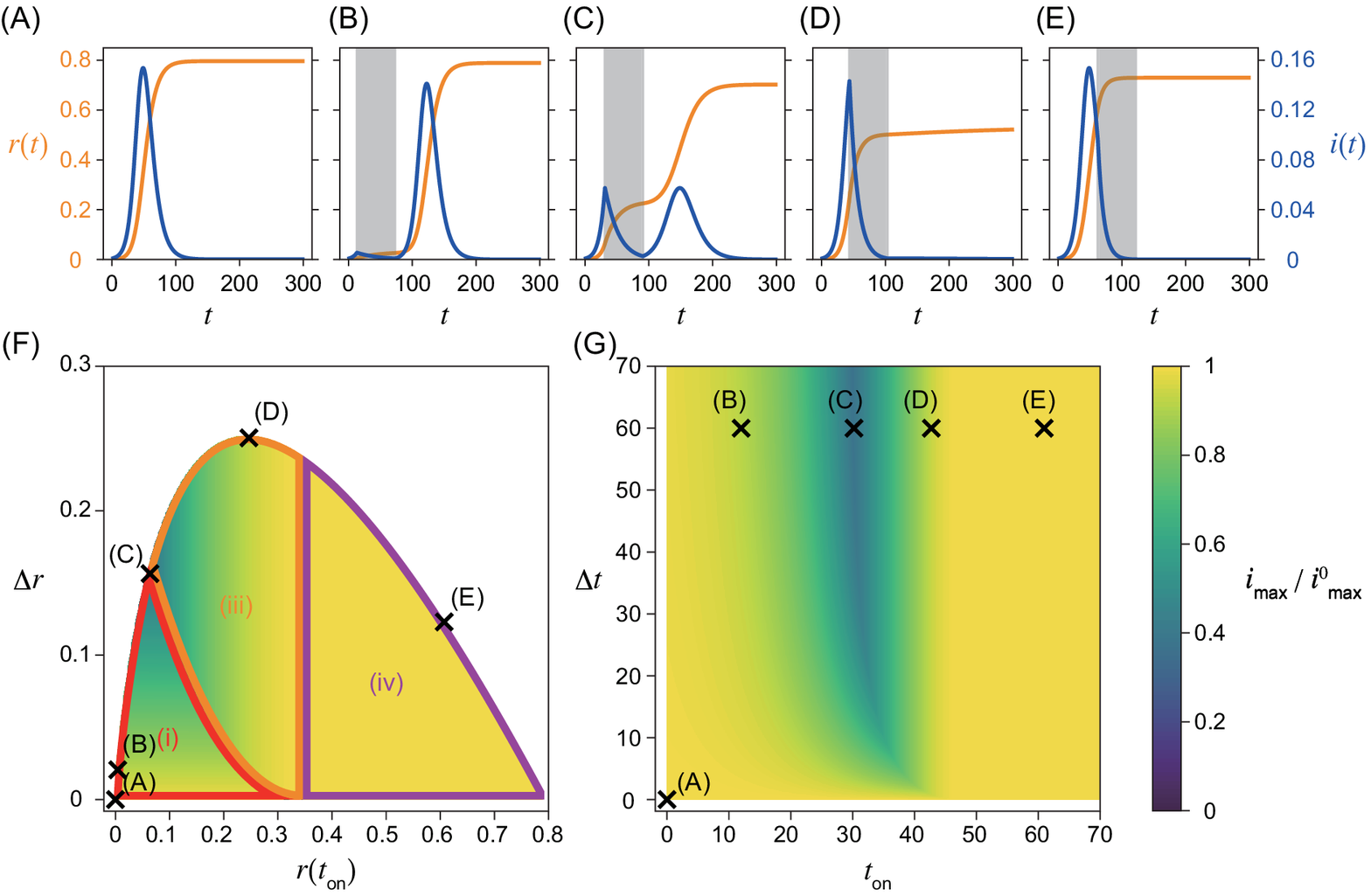}
		\caption{ Strong intervention with small $R_{0,\rm on}=0.7<1$. Parameters are $\beta_{\rm off}=2/7 \ {\rm days}^{-1}$ and $\gamma=1/7 \ {\rm days}^{-1}$. (A) Time series of the fraction of infected and removed individuals, $i(t)$ in blue and $r(t)$ in orange, without intervention. Note that this time series is identical to that presented in Fig.~\ref{fig:large_beta1}(A). Time series with  intervention with $\beta_{\rm on}=0.7/7 \ {\rm days}^{-1}$ and the intervention duration $\Delta t=60$ days with  onset times (B) $t_{\rm on}=12$ days, (C) $t_{\rm on}=30.2$ days, (D) $t_{\rm on}=42.7$ days, and (E) $t_{\rm on}=61$ days. The intervention is implemented for $t_{\rm on}\le t \le t_{\rm off}=t_{\rm on}+\Delta t$ and are represented by the grey intervals. 
			%Panels (F) and (G) and panels (H) and (I) represent the maximum of the fraction of the infected individuals $i_{\rm max}$ normalized by that without the intervention $i_{\rm max}^0$ and the final size $r(\infty)$ normalized by that without the intervention $r^0(\infty)$, respectively. Panels (F) and (H) and panels (G) and (I) are plotted in terms of $r(t_{\rm on})$ and $\Delta r$, and $t_{\rm on}$ and $\Delta t$, respectively. Symbols (A)-(E) in these panels denote the parameter values corresponding to the time series in (A)-(E).
			Symbols (i), (ii), and (iv) in panel (F) denote the times at which the maximum infected fraction is observed. Region (ii) is not observed in this case. Notations of other symbols are the same as those in Fig.~\ref{fig:large_beta1}.
		}
		%    \label{fig:small_beta1}
		\label{fig:small_beta1}
	\end{figure}

	Equations (\ref{eq:s}) and (\ref{eq:i}) were numerically integrated using ODEPACK \cite{h83}, which consists of nine Fortran solvers for the initial value problem for ODE systems. The time step was set to $0.001$. 
	%We varied intervention transmission rate $\beta_{\rm on}$, the initial intervention time $t_{\rm on}$, and the intervention duration $\Delta t$.
	The nonintervention transmission rate and the recovery rate are fixed as $\beta_{\rm off}=2/7 \ {\rm days}^{-1}$ and $\gamma=1/7 \ {\rm day}^{-1}$, that is, $R_{0,\rm off}=2$.   The initial condition $s(0)=1-\epsilon$, $i(0)=\epsilon$, and $r(0)=0$, where $\epsilon=0.001$ is employed in this study. As discussed in detail below, the system behaves qualitatively differently for different values of $R_{0,\rm on}$. Here, we primarily focus on the cases for $R_{0,\rm on}=1.4$ (Fig.~\ref{fig:large_beta1}) and $R_{0,\rm on}=0.7$ (Fig.~\ref{fig:small_beta1}), corresponding to relatively weaker and stronger intervention intensities, respectively. Note that $R_{0,\rm on}>1$ in the former case implies that the infection may spread even in the presence of the intervention. The time series without the intervention is shown in Figs.~\ref{fig:large_beta1}(A) and \ref{fig:small_beta1}(A). They are identical, because $R_{0,\rm on}$ does not affect the dynamics without the intervention. Figures \ref{fig:large_beta1}(B)--(E) and \ref{fig:small_beta1}(B)--(E) show the time series with an intervention with a constant intervention duration $\Delta t=60$ days. The second wave can occur if the herd immunity is not achieved when the intervention ends (Figs.~\ref{fig:large_beta1}(C) and \ref{fig:small_beta1}(B), (C)).

	% For figure citations, please use "Fig" instead of "Figure".
	
	\subsection*{Maximum fraction of infected individuals}
	In addition to the numerically obtained time series, we analytically show that peaks of the fraction of infected individuals can appear at the following four timings by applying Eq.~(\ref{eq:s0_t}): after the intervention, during the intervention, at the onset of the intervention, and before the intervention. Note that there can be two peaks of the fraction of infected individuals if a second wave occurs. In such a case, the peak with a larger fraction gives the global maximum. Let us describe the four cases with respect to the timing of the maximum in detail. 
	
	\begin{itemize}
		\item (i) The maximum appears after the intervention (Figs.~\ref{fig:large_beta1}(B), (C), \ref{fig:small_beta1}(B), and (C)). If the intervention ends before achieving herd immunity, a peak is observed during the second wave after the intervention. The fraction of infected individuals for this peak gives the global maximum if it is higher than the first peak before or during the intervention. 
		
		\item (ii) The maximum appears during the intervention (Fig.~\ref{fig:large_beta1}(D)). If the effective reproduction number $R_t$ declines and crosses unity during the intervention, there is a peak in this timing. The condition $R_{0, \rm on}>1$ is necessary for the existence of this peak, because the effective reproduction number has to be larger than unity at the onset of the intervention. Therefore, this peak does not appear for $R_{0,\rm on}<1$.
		This peak is the global maximum if it is larger than the second peak. 
		The peak also appears at this timing in Fig.~\ref{fig:large_beta1}(C), but the second peak is slightly higher than this peak.
		
		\item (iii) The maximum appears at the onset of the intervention at $t=t_{\rm on}$  (Fig.~\ref{fig:small_beta1}(D)). If $R_{t_{\rm on}}<1$ holds when the intervention starts, the fraction of infected individuals decreases during the intervention, and we observe a local peak at $t=t_{\rm on}$. There may be another peak if herd immunity is not achieved during the intervention. If the fraction of infected individuals at the second peak is less than at this peak, the timing of the global maximum is given at $t_{\rm on}$.
		%\textcolor{red}{CHECK Figure \ref{fig:small_beta1}(C) shows an example where this peak gives the global maximum.}
		Note that this region  also appears for $R_{0,\rm on}>1$, although the time series is not shown in Figs.~\ref{fig:large_beta1}(B)--(E).

		\item (iv) The maximum appears before the onset of the intervention (Figs.~\ref{fig:large_beta1}(E) and \ref{fig:small_beta1}(E)).  The fraction of infected individuals reaches its maximum before the intervention. This case implies that the intervention starts too late and fails to mitigate outbreaks in terms of the maximum fraction of infected individuals.
		% \begin{align}
		%     i(t_{p_1})     &= 1 - \frac{\gamma}{\beta_{\rm off}} \bigg[ 1 + \log \left( \frac{\beta_{\rm off}}{\gamma} \right) \bigg]
		% \end{align}

		% \begin{align}
		%     i(t_{\rm on})     &= 1 - e^{-\frac{\beta_{\rm off}}{\gamma}r(t_{\rm on})} - r(t_{\rm on})
		% \end{align}

		% \begin{align}
		%     i(t_{p_2})     &=
		%      \frac{\beta_{\rm off} - \beta_{\rm on}}{\beta_{\rm on}} r(t_{\rm on})
		%      +
		%      1-\frac{\gamma}{\beta_{\rm on}} \bigg[ 1+\log \left(\frac{\beta_{\rm on}}{\gamma} \right) \bigg]. 
		% \end{align}

		% \begin{align}
		%     i(t_{p_3}) &= \frac{\beta_{\rm off} - \beta_{\rm on}}{\beta_{\rm off}} [r(t_{\rm on})-r(t_{\rm off})] + 1 - \frac{\gamma}{\beta_{\rm off}} \bigg[ 1 + \log \left(\frac{\beta_{\rm off}}{\gamma} \right)  \bigg]. 
		% \end{align}
	\end{itemize}
	%Typical time series of these four cases 
	%obtained by numerically integrating Eqs.~(\ref{eq:s}), (\ref{eq:i}) 
	%are shown in Fig.~\ref{fig:small_beta1}. 
	%For cases B, D1, and D2, there are two local maxima in the fraction of the infected individuals.
	The fraction of infected individuals at the peaks can be calculated analytically. Conditions for the peaks are given in terms of the fraction of susceptible individuals as $s(t)=1/R_{0,{\rm off}}=\gamma/\beta_{\rm off}$ without the intervention and $s(t)=1/R_{0,{\rm on}}=\gamma/\beta_{\rm on}$ with the intervention.
	The maximum fraction of infected individuals for cases (ii), (iii), and (iv) does not depend on the fraction of removed individuals at the offset of the intervention $r(t_{\rm off})$, because these peaks appear before $t_{\rm off}$.
	For case (i), the maximum fraction of infected individuals $i_{\rm max}$ is given as
	\begin{eqnarray}
		i_{\rm max} &=& 1 - \left( 1- \frac{R_{0,\rm on}}{R_{0,\rm off}}\right)  \Delta r
		%[r(t_{\rm on})-r(t_{\rm off})] 
		- \frac{1}{R_{0,\rm off}} \bigg[ 1 + \log \left(R_{0,\rm off} \right)  \bigg],
		%. \\
		%i_{\rm max} &= \frac{R_{0,\rm off} - R_{0,\rm on}}{R_{0,\rm off}} \big[r(t_{\rm on})-r(t_{\rm off}) \big] + 1 - \frac{1}{R_{0,\rm off}} \bigg[ 1 + \log \left(R_{0,\rm off} \right)  \bigg], 
		%\\
		%&= \textcolor{red}{DELETE \frac{\beta_{\rm off} - \beta_{\rm on}}{\beta_{\rm off}} [r(t_{\rm on})-r(t_{\rm off})] + 1 - \frac{\gamma}{\beta_{\rm off}} \bigg[ 1 + \log \left(\frac{\beta_{\rm off}}{\gamma} \right)  \bigg], }
		\label{eq:peak_d}
	\end{eqnarray}
	which depends on $\Delta r:= r(t_{\rm off})-r(t_{\rm on})$, the difference in the fraction of removed individuals between the onset and offset of the intervention.
	See Supplementary Section 2, Additional File 1 for the explicit form of $i_{\rm max}$ for all cases and its derivation.
	
	The boundaries between regions (i) and (ii), (i) and (iii), (ii) and (iii), and (iii) and (iv) can be obtained analytically with respect to $r(t_{\rm on})$ and $\Delta r$ (Figs.~\ref{fig:large_beta1}(F) and \ref{fig:small_beta1}(F)). On the boundaries, the fraction of infected individuals at two peaks is comparable (Figs.~\ref{fig:large_beta1}(C) and \ref{fig:small_beta1}(C)). See Supplementary Section 3, Additional File 1 for details of the derivation.
	
	\subsection*{Final size with intervention}
	
	The final size also reflects the effect of the intervention. 
	The final size of removed individuals $r(\infty)$ with the intervention is obtained by solving the equation 
	\begin{eqnarray}
		r(\infty)
		&=& 1 - \exp \bigg[\big(R_{0,\rm off}-R_{0,\rm on}\big) \Delta r \bigg] \exp \bigg[- R_{0,\rm off} \ r(\infty) \bigg], \label{eq:final} 
		%   &= 1 - \exp \bigg\{\big(R_{0,\rm off}-R_{0,\rm on}\big) \big[r(t_{\rm off})-r(t_{\rm on})\big] \bigg\} \exp \bigg[- \frac{\beta_{\rm off}}{\gamma} r(\infty) \bigg], \label{eq:final} 
	\end{eqnarray}
	in a self-consistent manner \cite{fow20}.
	Specifically, as $r(\infty)$ appears on both sides, this equation can be solved numerically or using the Lambert $W$ function, except for some special cases.
	This equation implies that the final size depends on $\Delta r = r(t_{\rm off})-r(t_{\rm on})$.
	Another important implication of this equation is that $r(t_{\rm off})$ has the upper bound $\tilde{r}$ depending on $r(t_{\rm on})$, which is given by
	\begin{eqnarray}
		\tilde{r}(r(t_{\rm on})) &=& 1 - \exp \bigg[-(R_{0,\rm off} - R_{0,\rm on}) r(t_{\rm on})\bigg] \exp\bigg[-R_{0,\rm on} \   \tilde{r}\bigg].
		%    \tilde{r}(r(t_{\rm on})) &= 1 - \exp \left[-\frac{\beta_{\rm off} - \beta_{\rm on}}{\gamma} r(t_{\rm on})\right] \exp\left[-\frac{\beta_{\rm on}}{\gamma}  \tilde{r}\right].
		\label{eq:final_intervention}
	\end{eqnarray}
	See Supplementary Section 4, Additional File 1 for the details of the derivation.
	%the difference between the fraction of the removed individual at the onset of the intervention $r(t_{\rm on})$ and that at the offset of the intervention $r(t_{\rm off})$.
	
	Using this equality, one can show that the final size in the presence of the intervention is always smaller than that without the intervention.
	For
	\begin{eqnarray}
		R_{0,\rm on}  &< &\frac{R_{0,\rm off} }{R_{0,\rm off}-1}  \log \left(R_{0,\rm off} \right) ,
		\label{eq:opt_timing}
	\end{eqnarray}
	one can achieve $r(\infty)\approx 1-1/R_{0,\rm off}$ by setting $t_{\rm on}$ properly, which is the smallest prevalence to achieve herd immunity, with an intervention duration $\Delta t$ that is large enough \cite{fow20}.
	See Supplementary Section 5, Additional File 1 for details of the derivation.
	Numerical results regarding the final size are summarized in Supplementary Section 6, Additional File 1.

	\section*{Results}
	%\subsection*{Time series}
	We report the numerical and analytical results, when the reproduction number under the intervention $R_{0,\rm on}$ is large (Fig.~\ref{fig:large_beta1}) and small (Fig.~\ref{fig:small_beta1}), showing qualitatively different behaviors.
	
	In Figs.~\ref{fig:large_beta1}(F), (G), \ref{fig:small_beta1}(F), and (G), the dependence of the maximum fraction of infected individuals on the timing of the intervention is plotted. Here,  $i_{\rm max}$ is normalized by that in the absence of  the intervention  $i_{\rm max}^{0}$ (Figs.~\ref{fig:large_beta1}(A) and \ref{fig:small_beta1}(A)). As the maximum infected fraction drops in the presence of the intervention, $i_{\rm max}/i_{\rm max}^0$ is less than unity and quantifies the effectiveness of the intervention in terms of the maximum fraction of infected individuals. As this ratio decreases, the intervention shows more success in reducing the maximum fraction. 
	
	It is difficult to obtain the time series of the SIR model analytically without any approximations, for example, linearization, or the method presented in \cite{km27}. 
	Therefore, $i_{\rm max}$ is numerically computed with different onset and offset times for the intervention, $t_{\rm on}$ and $t_{\rm off}$ (Figs.~\ref{fig:large_beta1}(G) and \ref{fig:small_beta1}(G)).
	However, $i_{\rm max}$ can be analytically calculated with respect to the fraction of the recovered individuals at the onset, $r(t_{\rm on})$, and the offset, $r(t_{\rm off})$, of the intervention (Figs.~\ref{fig:large_beta1}(F) and \ref{fig:small_beta1}(F)). 
	Note that there exists a one-to-one correspondence between $\{t_{\rm on},t_{\rm off} \}$ (panels (F)) and $\{ r(t_{\rm on}), r(t_{\rm off}) \}$ (panels (G)) in Figs.~\ref{fig:large_beta1} and \ref{fig:small_beta1}.

	%We discuss two different $R_{0,\rm on}$ values in the following. 
	
	\subsection*{Weak intervention (large $R_{0,\rm on}=1.4$) }
	In this case, the reproduction number is larger than unity even in the presence of the intervention. Therefore, the fraction of infected individuals may increase during the intervention period. The peaks of infected individuals can be observed during the intervention, and the maximum infected fraction can appear at any of the four timings (i)--(iv) classified above.
	%The maximum fraction of the infected individuals is small with different $t_{\rm on}$ is given in Figs.~\ref{fig:large_beta1}(B)-(E). The timing giving the maximum is (i) after the intervention in panel (B), (ii) during the intervention in panels (C) and (D), and (iv) before the intervention in panel (E).
	Figures \ref{fig:large_beta1}(B)--(E) show that $i_{\rm max}$ is minimized in the intermediate onset time $t_{\rm on}$ near (C), where the peaks during and after the intervention are comparable. This non-monotonic dependence on $t_{\rm on}$ is clearly visualized in Fig.~\ref{fig:large_beta1}(G).
	%, where $i_{\rm max}/i_{\rm max}^0$ is plotted in $r(t_{\rm on})$-$\Delta r$ plane. if the intervention starts at $t_{\rm on}=0$ (Fig.~\ref{fig:large_beta1}(B)). 
	The peak of infected individuals during the intervention is smaller than that without intervention. This intervention mitigates the second wave.  
	
	%\subsubsection*{Maximum fraction of the infected individuals}
	As shown in Eq.~(\ref{eq:peak_d}) and Fig.~\ref{fig:large_beta1}(F), the maximum fraction of infected individuals is linear in $\Delta r$ if the peak of the second wave is the maximum, that is, case (i). This is verified by the fact that the contours lie horizontally in case (i). To clarify this point, the contours are explicitly shown in Supplementary Section 7, Additional File 1. 
	If the fraction of the infected individuals reaches the maximum during or at the onset of the intervention (cases (ii) and (iii), respectively), the maximum fraction depends only on $t_{\rm on}$, which has one-to-one correspondence to $r(t_{\rm on})$. This is verified in Fig.~\ref{fig:large_beta1}(F), (G), and Supplementary Section 7, Additional File 1. 
	If the maximum appears before the intervention,  that is, case (iv), the maximum is independent of both $t_{\rm on}$ and $t_{\rm off}$. 
	
	Onset and offset times for the intervention with constant $\Delta t$ corresponding to Figs.~\ref{fig:large_beta1}(A)--(E) are plotted in Figs.~\ref{fig:large_beta1}(F) and (G). As suggested by the time series, the maximum infected fraction is smallest in the intermediate intervention onset $t_{\rm on}$, near (C).
	It is clear from panel (F) that this point is located close to the boundary between cases (i) and (ii), where peaks during and after the intervention are comparable. In case (ii), the maximum does not depend on $\Delta r$, which implies that a longer intervention does not reduce the maximum.
	Note that the relationship between $(t_{\rm on},t_{\rm off})$ and $(r(t_{\rm on}),r(t_{\rm off}))$ is non-monotonous. %\textcolor{red}{Non-monontonic }

	\subsection*{Strong intervention (small $R_{0,\rm on}=0.7$) }
	%\subsubsection*{Maximum fraction of the infected individuals}
	
	When the transmission rate is small during the intervention,  the maximum infected fraction is minimized in the intermediate starting time of the intervention $t_{\rm on}$, namely, early implementation of the intervention does not necessarily minimize the infected fraction (Figs.~\ref{fig:small_beta1}(F), (G)). 
	This result is intuitively understood as follows. If the intervention starts too early, the infection does not spread because of the small intervention transmission rate. Therefore, the second wave after the intervention is large, thus the early intervention is not effective. 
	If the timing of the intervention is characterized in terms of $r(t_{\rm on})$ and $r(t_{\rm off})$ (Fig.~\ref{fig:small_beta1}(F)), the maximum of the second wave depends only on $\Delta r$. 
	If the maximum fraction is found during the intervention, its value is independent of $t_{\rm off}$ and $r(t_{\rm off})$. The peak does not appear during the intervention, that is, case (ii) does not appear because $R_{t,\rm on}<1$ holds.

	\subsection*{Maximum infected fraction $\bar{i}_{\rm max}$ versus reproduction number under the intervention $R_{0,\rm on}$}
\begin{figure}
		\centering
		\includegraphics[width=0.75\textwidth]{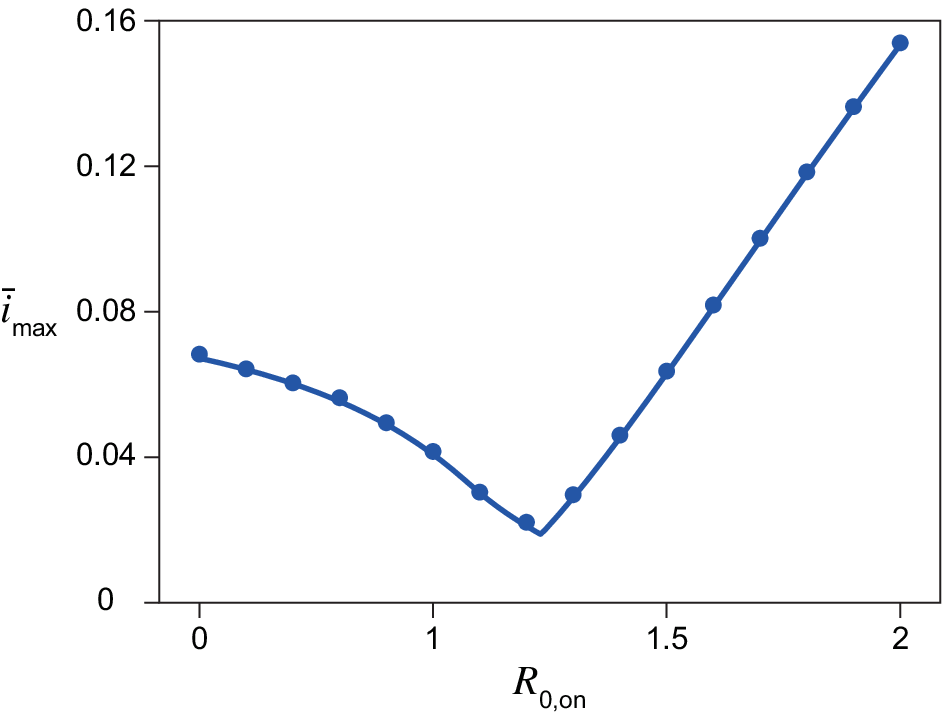}
		\caption{Dependence of $\bar{i}_{\rm max}$, the maximum fraction of infected individuals minimized by choosing the onset and offset timings of intervention, on the intervention  reproduction number $R_{0,\rm on}$. The symbols and solid line represent the numerical and analytical results (Supplementary Section 8, Additional File 1), respectively. The numerical results verifies the theoretical prediction that $\bar{i}_{\rm max}$ takes the minimum value at $R^\ast_{0,\rm on}\approx 1.23$. 
		}
		\label{fig:peak_beta1}
	\end{figure}

	For each $R_{0,\rm on}$, there exist onset and offset timings of the intervention that minimize the maximum fraction of infected individuals $i_{\rm max}$ (Figs.~\ref{fig:large_beta1}(F), (G) and \ref{fig:small_beta1}(F), (G)). Let this value  be $\bar{i}_{\rm max}(R_{0,\rm on})$.
	Figure \ref{fig:peak_beta1} plots the dependence of $\bar{i}_{\rm max}$ on $R_{0,\rm on}$. As seen in the figure, it is minimized at a non-trivial intermediate value of $R_{0,\rm on}=R_{0,\rm on}^{\ast}\approx 1.23>1$. This implies that the maximum fraction of infected individuals is minimized for a weak intervention under the one-shot condition. 
	For  $R_{0,\rm on}\ge R_{0,\rm on}^\ast$, $\bar{i}_{\rm max}$ is achieved at the boundary between regions (i) and (ii) at $t_{\rm on}=0$, where the peaks during and after the intervention are comparable (Fig.~\ref{fig:large_beta1}(C)). For $R_{0,\rm on}^\ast \ge R_{0,\rm on} \ge R_{0,\rm on}^{(1)}$, the boundary between regions (i) and (ii) with $t_{\rm on}>0$ gives $\bar{i}_{\rm max}$, where $R_{0,\rm on}^{(1)}\approx 1.08$ is the parameter value below which region (ii) does not exist. For strong intervention $R_{0,\rm on}\le R_{0,\rm on}^{(1)}$, $\bar{i}_{\rm max}$ is found at the boundary between regions (i) and (iii), where the peaks of the onset and after the intervention are comparable (Fig.~\ref{fig:small_beta1}(C)), at $t_{\rm on}>0$.
	Conditions for $\bar{i}_{\rm max}$ are analytically derived in all cases, shown by the solid line in Fig.~\ref{fig:peak_beta1}. The conditions for $\bar{i}_{\rm max}$ for $R_{0,\rm on}\ge R_{0,\rm on}^{(1)}$ can be explicitly solved. The condition for small $R_{0,\rm on}\le R_{0,\rm on}^{(1)}$ cannot be solved explicitly, and the parametric equations for $\bar{i}_{\rm max}$ and $R_{0,\rm on}$ are used to plot the theoretical curve in Fig.~\ref{fig:peak_beta1}. See Supplementary Section 8, Additional File 1 for details of the derivation.
	
	\section*{Intervention strategies}
	It is possile to optimize the onset and offset timings of the NPI by minimizing an objective function under certain constraints. In the present framework, this can be formulated as an optimization problem in a hybrid nonlinear dynamical system. The optimal intervention strategy depends on the objective function. 
	Here, we discuss the following two simple scenarios to minimize $i_{\rm max}$. In general, more complex objective functions can be used for optimization. The following discussion may provide some intuition for considering such cases, taking into account the second wave. 
	%\textcolor{red}{In the real situation, we should note that $\beta_{\rm on}$ is not easy to control and the time delay between the introduction of the intervention and the report of the number of cases. }

	\subsection*{Minimizing $i_{\rm max}$ with a constraint in the intervention duration}
	Let us start with a case where the intervention duration $\Delta t$ is less than a certain value. As $i_{\rm max}$ is monotonically decreasing in $\Delta t$ for a fixed $t_{\rm on}$, we can assume that $\Delta t$ is a constant, and  $t_{\rm on}$ is varied.
	This case has been studied in Figs.~\ref{fig:large_beta1}(B)--(E) and \ref{fig:small_beta1}(B)--(E). This scheme fixes the intervention duration, so it is easier to anticipate the economic impact of the intervention, which depends on the duration of the intervention, than in the next scenario based on the fraction of  recovered individuals. 
	
	As discussed above, $i_{\rm max}$ is minimized in the intermediate onset time $t_{\rm on}$, when the first peak during the intervention and the second peak after the intervention are comparable. For larger $R_{0,\rm on}$, $t_{\rm on}$ giving the minimum $i_{\rm max}$ converges to zero for large $\Delta t$ (Fig.~\ref{fig:large_beta1}(G)), but converges to non-zero for large $\Delta t$ for smaller $R_{0,\rm on}$ (Fig.~\ref{fig:small_beta1}(G)). These results are understood as follows: for a strong intervention with small $R_{0,\rm on}$, the intervention immediately suppresses the fraction of infected individuals. Therefore, the early onset of the intervention prevents more individuals from achieving immunity during the intervention and eventually increases the maximum fraction of infected individuals. 
	
	\subsection*{Minimizing intervention duration with a constraint in $i_{\rm max}$}
	Another possible constraint is to minimize the intervention duration $\Delta t$, keeping the maximum fraction of infected individuals $i_{\rm max}$ constant. This corresponds to choosing $t_{\rm on}$ along a contour of $i_{\rm max}$, which prevents the overcapacity of medical support. In Figs.~\ref{fig:large_beta1}(G) and \ref{fig:small_beta1}(G), $\Delta t$ is minimized for an intermediate $t_{\rm on}$. Namely, a contour that crosses point (C) ($\Delta t=60$) reaches $\Delta t\approx 20$ if the onset time is later than $t_{\rm on}=19.2$ (Fig.~\ref{fig:large_beta1}(C)).
	As we have already discussed, the maximum fraction depends on $\Delta r$ in region (i). It takes a shorter time to achieve the same $\Delta r$ if the intervention starts later in this case. 
	As the maximum fraction of infected individuals depends on $r(t_{\rm on})$ but is independent of $\Delta r$ in cases (ii) and (iii), the optimal $r(t_{\rm on})$ is determined by the boundary between regions (i) and (ii) or that between (i) and (iii) in Figs.~\ref{fig:large_beta1}(F) and \ref{fig:small_beta1}(F). Evidently this timing is an intermediate value of $t_{\rm on}> 0$.
	%\textcolor{red}{It is worth emphasizing again that the global minimum is reached for nontrivial $R_{\rm on}$ (Fig.~\ref{fig:peak_beta1}).}
	It may be easier to set a plan for the intervention with respect to $r(t_{\rm on})$ and $r(t_{\rm off})$, rather than the timing $t_{\rm on}$ and $t_{\rm off}$.
	
	Similar optimization problems can be considered using the final size as the objective functions. These problems are discussed in Supplementary Section 9, Additional File 1.
	
	\section*{Conclusion}
	In the present study, in consideration of the actual COVID-19 situation, we studied the situation in which the reproduction number of an infectious disease is temporarily reduced by implementing an NPI once during the epidemic, using a simple mathematical model. The results provide theoretical implications as to how strong NPIs should be introduced during an epidemic of an emerging infectious disease. If the effective reproduction number during the intervention is too small, the fraction of infected individuals at the peak in the second wave may be higher than the first peak. It was also shown numerically and analytically that the fraction of infected individuals can also increase if the intervention is started too early. The upper limit of medical capacity is an essential practical constraint. In particular, for infectious diseases such as COVID-19, which is too emergent to expect effective treatments, it is more important to avoid exhausting the medical system. This study suggests that it will be necessary to be alert for a larger second wave that may occur after strong intervention in such cases.
	
	We analytically derived the peak fractions of infected individuals in the SIR model with the one-shot intervention. These analytical results suggest that the peak fractions can be smaller with non-trivial intervention timings. The maximum fraction is smallest for $R_{0,\rm on}>1$, that is, the intervention reproduction number is not in the disease-free regime.  In the literature, Bootsma and Ferguson \cite{bf07} showed that the final size can be minimized for non-trivial intervention timing and $R_{0,\rm on}$ numerically. Di Lauro et al. \cite{dkm20} numerically showed that the peak fraction of infected individuals also depends on the timing of the intervention. We obtained analytical expressions for these quantities in this study. The analytical results  for the final size are presented in Supplementary Section 6, Additional File 1. It should be noted that the formulae for the peak fraction depend on the timing of the peak, resulting in various cases compared with the final size. Some analytical results are available for this system. Sadeghi et al. \cite{sgs20} explained these non-trivial effects based on solutions of the linearized equation, which exponentially grows and decays without and with the intervention, respectively. Linearization is one of the simplest approximations and is applicable in this case; in particular, this approximation is useful in  discussions regarding timing. Conversely, 
	%, these approximations may be inconsistent with the numerical results. In particular, 
	linearization cannot be used to discuss the important case where $R_{0,\rm on}>1$, as the linearized equation cannot explain the declining number of infected individuals. The proposed method in this study provides a unified framework, including the cases where linearization is not feasible. Morris et al. derived \cite{mrp20} an equation for the peaks of the fraction of infected individuals in terms of $s(t_{\rm on})$, $s(t_{\rm off})$, $i(t_{\rm on})$, and $i(t_{\rm off})$. In this work, we further show the peak fractions in terms of the two parameters $r(t_{\rm on})$ and $r(t_{\rm off})$ for a general initial condition.

	%Whereas concrete measures and guidelines for COVID-19 are required, it is emphasized again that this study is merely a simplified model based on many assumptions. First, the results presented in this study are based on the simplified SIR model which does not take into account the realistic pathology of COVID-19. This model assumes uniform and random contact within a group and does not consider interactions between different subgroups in the population. Recovered patients are assumed to have complete immunity in the model. These assumptions are not applicable to COVID-19, where there are still many unclear points regarding heterogeneity in contact networks and immune response of patients after recovery. Next, this study is a one-shot intervention model, in which the transmission coefficient returns to the original value after single intervention. Practically, each re-pandemic requires multiple intermittent interventions \cite{fni20-9}, making the intervention process much more complex \cite{ktg20}. These complex situations may be analyzed in detail by extending the methods presented in this study.
	Although concrete measures and guidelines for COVID-19 are required, it is emphasized again that the results in this study were derived using a simplified model with many assumptions. First, the results presented in this study are based on the simplified SIR model, which takes into account neither the realistic pathology of COVID-19 nor societal response.  This model assumes uniform and random contact within a group and does not consider interactions between different subgroups in the population. Recovered patients are assumed to have complete immunity in this model. These assumptions are not applicable to COVID-19, where there are still many unclear factors regarding heterogeneity in contact networks and the immune response of patients after recovery. If a society develops other public health measures during the intervention, the basic reproduction numbers before and after the NPI may be different. Next, this study is a one-shot intervention model, in which the transmission coefficient returns to the original value after a single intervention. Practically, each re-pandemic requires multiple intermittent interventions \cite{fni20-9}, making the intervention process much more complex \cite{ktg20}. Furthermore, the model is based on a deterministic dynamical system of an isolated population. Therefore, important NPI measures such as border control cannot be estimated in the present framework. The deterministic nature of the model assumes that the infectious disease cannot be eradicated, as the number of infected individuals remains non-zero for a finite time. If the population is small enough and no imported cases are assumed, strong and early intervention, which is not necessarily recommended in this study, may eradicate the disease, and a second wave does not occur. Strong intervention would be necessary in other cases, for example, when the number of infected patients approaches the capacity of the medical system. These complex situations may be analyzed in detail by extending the methods presented in this study, which may lead to different conclusions from those reached in this study.
	
	In this study, the effect of NPIs was modeled as  a hybrid dynamical system, which may further enable us to approach more refined models in future investigations. The influence of NPIs with respect to political decisions and behavioral changes of people can be expressed more accurately by introducing a hybrid dynamical system. In recent years, dynamical systems theory and control theory have been developed, and phenomena specific to hybrid systems such as Zeno solutions and sliding motions have been discussed \cite{f88,ll09,as10,dh10,hdl10}. Some studies, such as \cite{xxt12} and \cite{khz16}, have proposed control of infectious diseases using sliding mode control. The optimal policy under NPIs can be discussed by modeling the effect of economic damages associated with execution and then minimizing the cost function. Discrete changes in parameters such as the transmission coefficient in NPI implementation are strongly linked to the intensity of measures, suppression of economic activity, and changes in human mobility. For example, the correlation between the decrease in human mobility with NPIs and the effective reproduction number for the COVID-19 pandemic has been studied  \cite{bdm20,ytf20}. Such studies can  contribute to modeling the costs of NPIs. The mathematical model of the epidemic suppression effect can be constructed using a hybrid dynamical system, taking into account the negative socioeconomic impact of NPIs.
	
	%\textcolor{red}{MOVE TO DISCUSSION Hybrid dynamics implemented in this study may also provide a framework which estimates the effects of NPIs taking into account socioeconomic factors which affect the transmissibility of the infection.}

	%%%%%%%%%%%%%%%%%%%%%%%%%%%%%%%%%%%%%%%%%%%%%%
	%%                                          %%
	%% Backmatter begins here                   %%
	%%                                          %%
	%%%%%%%%%%%%%%%%%%%%%%%%%%%%%%%%%%%%%%%%%%%%%%
	
	\begin{backmatter}
		
		\section*{Competing interests}
		The authors declare that they have no competing interests.
		
		\section*{Author's contributions}
		N.F., T.W., and K.A. designed the study. N.F. derived the analytical results. T.O., S.T., and K.A. provided comments from theoretical viewpoints. T.O. conducted the numerical simulations. T.W., J.S., and T.N. discussed the implications of the study with respect to public health.
		
		\section*{Acknowledgements}
		K.A. and N.F. are grateful to Prof. H. Inaba for their valuable comments. T.N. and N.F. are supported by Starting Grants for Research toward Resilient Society (SGRRS), Tohoku University. N.F. is supported by JSPS KAKENHI Grant Number JP18K11462. K.A. is partially supported by Moonshot R\&D Grant Number JPMJMS2021.

		\section*{Additional Files}
		\subsection*{Additional file 1 --- Supplementary texts}
		Detailed discussions carried out in the main text are given.
		
		%   \subsection*{Additional file 2 --- Sample additional file title}
		%     Additional file descriptions text.

	\end{backmatter}

\clearpage
\renewcommand{\figurename}{Figure S}
\renewcommand{\theequation}{S\arabic{equation}}

\section*{Supplementary materials: Analytical estimation of maximum fraction of infected individuals with one-shot non-pharmaceutical intervention in a hybrid epidemic model}

\section*{S1 Method: Derivation of Eq.~(4) \cite{km27,am92,cb09,i17}}
We want to derive Eq.~(4) in the main text from the evolution equation of the SIR model
\begin{eqnarray}
	\frac{ds}{dt} &=& -\beta s i, \\ %\label{eq:s}\\
	\frac{di}{dt} &=& \beta s i - \gamma i, \\ %\label{eq:i} \\
	\frac{dr}{dt} &=& \gamma i. \\ %\label{eq:r}
\end{eqnarray}
We integrate Eq.~(\ref{eq:s}) from time $t_0$ to $t_1$:
\begin{eqnarray}
	s(t_1) &=& s(t_0) \exp \left[-\beta \int_{t_0}^{t_1} i(t') dt'\right].
\end{eqnarray}
The integral on the right-hand side can be replaced with the integration of Eq.~(\ref{eq:r}), 
\begin{eqnarray}
	r(t_1) - r(t_0) &=& \gamma \int_{t_0}^{t_1} i(t') dt'.
\end{eqnarray}
Then, we obtain
\begin{eqnarray}
	s(t_1) &=& s(t_0) \exp\bigg\{ - \frac{\beta}{\gamma} [r(t_1)-r(t_0)] \bigg\}, \\ %\label{eq:s0_t}
\end{eqnarray}
which is Eq.~(4) in the main text. We apply this equation to various $t_0$ and $t_1$.
%This equation has been derived in \cite{fow20}.

\section*{S2 Method: Analytical expressions for the maximum fraction of infected individuals}
As described in the main text, there are four possible timings where the maximum fraction of infected individuals $i_{\rm max}$ is observed. We summarize the analytical forms of the maximum fraction in all cases with respect to the fractions of removed individuals at the onset $r(t_{\rm on})$ and the offset $r(t_{\rm off})$ of the intervention, that is, we assume that the timing of the onset and offset of the intervention is determined by the fraction of removed individuals.

\begin{itemize}
	\item \textbf{(i) The maximum appears after the intervention:}
	\begin{eqnarray}
		i_{\rm max} &=&
		1 - \left( 1- \frac{R_{0,\rm on}}{R_{0,\rm off}}\right)  \Delta r  - \frac{1}{R_{0,\rm off}} \bigg[ 1 + \log \left(R_{0,\rm off} \right)  \bigg],
		%    1 - \left( 1- \frac{R_{0,\rm on}}{R_{0,\rm off}}\right)  [r(t_{\rm off})-r(t_{\rm on})]  - \frac{1}{R_{0,\rm off}} \bigg[ 1 + \log \left(R_{0,\rm off} \right)  \bigg]. 
		\label{eq:peak1}
	\end{eqnarray}
	where $\Delta r=r(t_{\rm off})-r(t_{\rm on})$.
	
	\item \textbf{(ii) The maximum appears during the intervention:}
	\begin{eqnarray}
		i_{\rm max}     &=& 1+
		\left( \frac{R_{0,\rm off} }{R_{0,\rm on}} -1 \right)  r(t_{\rm on})
		-\frac{1}{R_{0,\rm on}} \bigg[ 1+\log \left(R_{0,\rm on} \right) \bigg]. 
		\label{eq:peak2}
	\end{eqnarray}
	
	\item \textbf{(iii) The maximum appears at the onset of the intervention:} 
	\begin{eqnarray}
		i_{\rm max}     &=& 1 - \exp \left[ -R_{0,\rm off}\ r(t_{\rm on})\right]  - r(t_{\rm on}).
		\label{eq:peak3}
	\end{eqnarray}
	
	\item \textbf{(iv) The maximum appears before the onset of the intervention:}
	\begin{eqnarray}
		i_{\rm max}     &=& 1 - \frac{1}{R_{0,\rm off}} \bigg[ 1 + \log \left( R_{0,\rm off} \right) \bigg].
		\label{eq:peak4}
	\end{eqnarray}
\end{itemize}

\subsection*{Derivation}
Here, we derive Eqs.~(\ref{eq:peak1})--(\ref{eq:peak4}).  The conditions for the fraction of infected individuals to be the local peak are given by $s(t) = \gamma / \beta_{\rm off}=1/R_{0,\rm off}$ without the intervention and $s(t)=\gamma / \beta_{\rm on}=1/R_{0,\rm on} $ with the intervention.

First, let us apply Eq.~(\ref{eq:s0_t}) until the onset time of the intervention, with $\beta=\beta_{\rm off}$, $t_1=t_{\rm on}$, and $t_0 = 0$. Using the initial condition  $s(0)=1-\epsilon$, $i(0)=\epsilon$, $r(0)=0$, we obtain
\begin{eqnarray}
	s(t_{\rm on}) &=&  (1-\epsilon) \exp\bigg[ - \frac{\beta_{\rm off}}{\gamma} r(t_{\rm on}) \bigg].
	\label{eq:peak_on_s}
\end{eqnarray}
In the following, we take the limit $\epsilon\rightarrow 0$.

Substituting Eq.~(\ref{eq:peak_on_s}) into Eq.~(\ref{eq:s0_t}), one can derive the fraction of susceptible and removed individuals during and after the intervention.

\subsubsection*{(i) Maximum fraction after the intervention}
We derive the fraction of infected individuals after the intervention. Let $t_p^{\rm after}>t_{\rm off}$ be the time at which the infected fraction reaches the peak after the intervention.
The relationship between $s(t_{\rm off})$ and $r(t_{\rm off})$ is derived by substituting Eq.~(\ref{eq:peak_on_s}) with $\epsilon\rightarrow 0$ into Eq.~(\ref{eq:s0_t}) with $t_0=t_{\rm on}$ and $t_1=t_{\rm off}$ as follows:
\begin{eqnarray}
	s(t_{\rm off}) 
	%&=& \exp\bigg[ - \frac{\beta_{\rm off}}{\gamma} r(t_{\rm on}) \bigg] \exp\bigg\{ - \frac{\beta_{\rm on}}{\gamma} \bigg[r(t_p^{\rm during})-r(t_{\rm on})\bigg] \bigg\} \\
	&=& \exp\bigg[ - \frac{\beta_{\rm on}}{\gamma} r(t_{\rm off}) \bigg] \exp \bigg[ - \frac{\beta_{\rm off}-\beta_{\rm on}}{\gamma} r(t_{\rm on})\bigg].  
\end{eqnarray}
Then, by applying Eq.~(\ref{eq:s0_t}) setting $t_0=t_{\rm off}$ and $t_1=t_p^{\rm after}$, we have
\begin{eqnarray}
	s(t_p^{\rm after}) &=& 
	\exp \bigg[ - \frac{\beta_{\rm off}}{\gamma} r(t_p^{\rm after}) \bigg]
	\exp\bigg\{  \frac{\beta_{\rm off}-\beta_{\rm on}}{\gamma} \bigg[ r(t_{\rm off})-r(t_{\rm on}) \bigg] \bigg\}.
	\label{eq:s_after}
\end{eqnarray}
The condition for the peak after the intervention is 
$s(t_p^{\rm after})=\gamma/\beta_{\rm off}$.
Substituting this condition into Eq.~(\ref{eq:s_after}), the fractions of removed and infected individuals are given by:
\begin{eqnarray}
	r(t_p^{\rm after}) &=& \frac{\beta_{\rm off} - \beta_{\rm on}}{\beta_{\rm off}} [r(t_{\rm off})-r(t_{\rm on})] + \frac{\gamma}{\beta_{\rm off}} \log \left(\frac{\beta_{\rm off}}{\gamma} \right)    ,\\
	i(t_p^{\rm after}) &=&   1 - s(t_p^{\rm after}) - r(t_p^{\rm after}) \\
	&=& 1 - \frac{\beta_{\rm off} - \beta_{\rm on}}{\beta_{\rm off}} [r(t_{\rm off})-r(t_{\rm on})]  - \frac{\gamma}{\beta_{\rm off}} \bigg[ 1 + \log \left(\frac{\beta_{\rm off}}{\gamma} \right)  \bigg], \label{eq:peak3b}
\end{eqnarray}
which is equivalent to Eq.~(\ref{eq:peak1}) with $R_{0,\rm off}=\beta_{\rm off}/\gamma$ and $R_{0,\rm on}=\beta_{\rm on}/\gamma$.
Equation (\ref{eq:peak3b}) gives $i_{\rm max}$ if the maximum appears after the intervention.
Note that the fraction of infected individuals depends only on the onset and offset timings as $r(t_{\rm off})-r(t_{\rm on})$. The later the onset or sooner the offset is, the higher the peak is.

\subsubsection*{(ii) Maximum fraction during the intervention}
If the effective reproduction number at the onset of the intervention is greater than unity, $\beta_{\rm on} s(t_{\rm on}) / \gamma>1$, and that at the offset is less than unity, $\beta_{\rm on} s(t_{\rm off}) / \gamma<1$, then there is a peak  during the intervention. 
Let $t_p^{\rm during}$ be the time at which the infected fraction reaches the peak, where $t_{\rm on} < t_p^{\rm during} < t_{\rm off}$. We obtain the relationship between susceptible and removed individuals during the intervention by applying Eq.~(\ref{eq:s0_t}) with $t_0=t_{\rm on}$ and $t_1=t_p^{\rm during}$ and using Eq.~(\ref{eq:peak_on_s}) as the relationship between $s(t_{\rm on})$ and $r(t_{\rm on})$:
\begin{eqnarray}
	s(t_p^{\rm during}) 
	%&=& \exp\bigg[ - \frac{\beta_{\rm off}}{\gamma} r(t_{\rm on}) \bigg] \exp\bigg\{ - \frac{\beta_{\rm on}}{\gamma} \bigg[r(t_p^{\rm during})-r(t_{\rm on})\bigg] \bigg\} \\
	&=& \exp\bigg[ - \frac{\beta_{\rm on}}{\gamma} r(t_p^{\rm during}) \bigg] \exp \bigg[ - \frac{\beta_{\rm off}-\beta_{\rm on}}{\gamma} r(t_{\rm on})\bigg].  \label{eq:t_p}
\end{eqnarray}
The condition for the peak during the intervention is 
$s(t_p^{\rm during})=\gamma/\beta_{\rm on}$.
By substituting this condition into Eq.~(\ref{eq:t_p}), we obtain
% \begin{eqnarray}
% \frac{\gamma}{\beta_{\rm on}} &=&     \exp\bigg[ - \frac{\beta_{\rm on}}{\gamma} r(t_p^{\rm during}) \bigg] \exp \bigg[ - \frac{\beta_{\rm off}-\beta_{\rm on}}{\gamma} r(t_{\rm on})\bigg].
% \end{eqnarray}
% 
\begin{eqnarray}
	r(t_p^{\rm during}) &=& \frac{\gamma}{\beta_{\rm on}} \bigg[ - \frac{1}{\gamma}(\beta_{\rm off} - \beta_{\rm on}) r(t_{\rm on}) + \log \left(\frac{\beta_{\rm on}}{\gamma} \right) \bigg].
\end{eqnarray}
Then, the fraction of infected individuals at this peak is 
\begin{eqnarray}
	i(t_p^{\rm during}) &=& 1 - s(t_p^{\rm during}) - r(t_p^{\rm during}) \\
	% &=& 1 - \frac{\gamma}{\beta_{\rm on}} \bigg[ 1 - \frac{1}{\gamma}(\beta_{\rm off} - \beta_{\rm on}) r(t_{\rm on}) + \log \left(\frac{\beta_{\rm on}}{\gamma} \right) \bigg] \\
	&=&
	1+\frac{\beta_{\rm off} - \beta_{\rm on}}{\beta_{\rm on}} r(t_{\rm on})
	-\frac{\gamma}{\beta_{\rm on}} \bigg[ 1+\log \left(\frac{\beta_{\rm on}}{\gamma} \right) \bigg],
	\label{eq:peak_int}
\end{eqnarray}
which is equivalent to Eq.~(\ref{eq:peak2}).
As $\beta_{\rm off}>\beta_{\rm on}$ and $\frac{\beta_{\rm off}-\beta_{\rm on}}{\beta_{\rm on}}>0$, the fraction of infected individuals at this peak linearly increases compared with the fraction of removed individuals at the onset of the intervention $r(t_{\rm on})$.
Equation (\ref{eq:peak_int}) is the condition for the maximum if this peak is higher than another peak.

The conditions for the existence of this peak are $\beta_{\rm on} s(t_{\rm on}) > \gamma$ and $\beta_{\rm on} s(t_{\rm off}) < \gamma$, which are interpreted as
\begin{eqnarray}
	r(t_{\rm on}) &<& \frac{\gamma}{\beta_{\rm off}} \log\left( \frac{\beta_{\rm on}}{\gamma} \right) , \label{eq:exist_on}\\
	r(t_{\rm off}) &>& \frac{\gamma}{\beta_{\rm on}} \bigg[ - \frac{1}{\gamma}(\beta_{\rm off} - \beta_{\rm on}) r(t_{\rm on}) + \log \left(\frac{\beta_{\rm on}}{\gamma} \right) \bigg].
	\label{eq:exist_off}
\end{eqnarray}
Therefore, $\beta_{\rm on} > \gamma$ is required for the existence of this peak.

\subsubsection*{(iii) Maximum fraction at the onset of the intervention}
At the onset of the intervention, the fraction of infected individuals is given as 
\begin{eqnarray}
	i(t_{\rm on}) &=& 1- s(t_{\rm on})-r(t_{\rm on}) \\
	&=& 1 - \exp\bigg[ - \frac{\beta_{\rm off}}{\gamma} r(t_{\rm on}) \bigg] - r(t_{\rm on}),
	\label{eq:peak_on}
\end{eqnarray}
by substituting Eq.~(\ref{eq:peak_on_s}).

This peak appears if the effective reproduction number before the intervention is larger than unity $\beta_{\rm off} s(t_{\rm on}) > \gamma$ and that after the onset of the intervention is less than unity $\beta_{\rm on} s(t_{\rm on}) < \gamma$. 
These conditions are summarized in terms of $r(t_{\rm on})$ as follows:
\begin{eqnarray}
	\max \left\{0, \frac{\gamma}{\beta_{\rm off}} \log \left( \frac{\beta_{\rm on}}{\gamma} \right) \right\}< r(t_{\rm on})<\frac{\gamma}{\beta_{\rm off}} \log \left( \frac{\beta_{\rm off}}{\gamma} \right),
	\label{eq:exist_iii}
\end{eqnarray}
by substituting the above conditions into  Eq.~(\ref{eq:peak_on_s}).

\subsubsection*{(iv) Maximum fraction before the intervention}
Let $t_1=t_p^{\rm before}$ be the time at which the peak appears before the onset of the intervention.
We apply Eq.~(\ref{eq:s0_t}) with $\beta=\beta_{\rm off}$, $t_0 = 0$, and $t_1=t_p^{\rm before}$. Then, we obtain
% \begin{eqnarray}
%     s(t_p^{\rm before}) 
%     &=& \exp\bigg[ - \frac{\beta_{\rm off}}{\gamma} r(t_p^{\rm before}) \bigg],
% \end{eqnarray}
% or equivalently
\begin{eqnarray}
	r(t_p^{\rm before}) &=&  \frac{\gamma}{\beta_{\rm off}} \log \left[ \frac{1}{s(t_p^{\rm before})} \right].
\end{eqnarray}
The peak condition $s(t_p^{\rm before})=\gamma /\beta_{\rm off}$ leads to
\begin{eqnarray}
	r(t_p^{\rm before}) &=&  \frac{\gamma}{\beta_{\rm off}} \log \left( \frac{\beta_{\rm off}}{\gamma} \right).
	\label{eq:exist_before}
\end{eqnarray}
Substituting this equation into the conservation of total population  $i(t)=1-s(t)-r(t)$, we obtain 
\begin{eqnarray}
	i(t_p^{\rm before}) &=& 1 - \frac{\gamma}{\beta_{\rm off}} \left[ 1+\log \left( \frac{\beta_{\rm off}}{\gamma} \right) \right].
\end{eqnarray}
This peak appears for $\displaystyle r(t_{\rm on})>\frac{\gamma}{\beta_{\rm off}} \log \left( \frac{\beta_{\rm off}}{\gamma} \right)$, meaning a late onset of the intervention.

\section*{S3 Method: Derivation of boundaries between regions of different maxima}
As shown in Figs.~1(F) and 2(F) in the main text, the timing giving the maximum fraction of infected individuals switches in the $(r(t_{\rm on}),\Delta r)$ plane, where $\Delta r=r(t_{\rm off}) - r(t_{\rm on})$. These figures suggest possible transitions between regions (i) and (ii), between (i) and (iii), between (ii) and (iii), and between (iii) and (iv) listed above. At the boundaries, the peaks of the different timings are expected to be equal. Based on this idea, we sketch the derivations of the equations for the boundaries between these timings.

\subsection*{Boundaries between regions (i) and (ii)}
%\textcolor{red}{WHERE TO PUT??This exists for $\beta_{\rm on}>\beta_{\rm on}^\ast $, where $\beta_{\rm on}^\ast \approx 1.07828$.}

In regions where peaks of infected individuals during and after the intervention coexist, the global maximum is given by
\begin{eqnarray}
	i_{\rm max} &=& 
	\max [ i(t_p^{\rm after})    ,i(t_p^{\rm during})] \\
	&=&
	\max \bigg\{1-
	\frac{\beta_{\rm off} - \beta_{\rm on}}{\beta_{\rm off}} [r(t_{\rm off})-r(t_{\rm on})]  - \frac{\gamma}{\beta_{\rm off}} \bigg[ 1 + \log \left(\frac{\beta_{\rm off}}{\gamma} \right)  \bigg], \nonumber \\
	&& 1+ \frac{\beta_{\rm off} - \beta_{\rm on}}{\beta_{\rm on}} r(t_{\rm on})
	-\frac{\gamma}{\beta_{\rm on}} \bigg[ 1+\log \left(\frac{\beta_{\rm on}}{\gamma} \right) \bigg]
	\bigg \} .
\end{eqnarray}
There occurs a transition of the timing giving the global maximum, and the condition at the boundary is $i(t_p^{\rm after})=i(t_p^{\rm during})$, that is,
\begin{eqnarray}
	\Delta r %&=& 
	%    &=& - \left( \frac{\beta_{\rm off}}{\beta_{\rm on}} - 1 \right) r(t_{\rm on}) + \frac{\beta_{\rm off}}{\beta_{\rm off}-\beta_{\rm on}}\frac{\gamma}{\beta_{\rm off} \beta_{\rm on}}\left[ \beta_{\rm off} + \beta_{\rm off} \log \left(\frac{\beta_{\rm on}}{\gamma}\right) - \beta_{\rm on} -\beta_{\rm on} \log \left( \frac{\beta_{\rm off}}{\gamma}\right) \right] \\
	&=&
	- \frac{\beta_{\rm off}}{\beta_{\rm on}}  r(t_{\rm on}) + \frac{\gamma}{ \beta_{\rm on}}\left[ 1 + \frac{\beta_{\rm off} \log \left(\frac{\beta_{\rm on}}{\gamma}\right) - \beta_{\rm on}  \log \left( \frac{\beta_{\rm off}}{\gamma}\right)}{\beta_{\rm off}-\beta_{\rm on}} \right]. 
	\label{eq:equi_int_after}
\end{eqnarray}
This condition gives the boundary between regions (i) and (ii) in Fig.~1(F) in the main text.
For $\Delta r$ smaller than this condition, the peak of the second wave is larger than that during the intervention.

\subsection*{Boundaries between (i) and (iii)}
Two peaks at the onset of the intervention and after the intervention can coexist. The maximum fraction of infected individuals is
\begin{eqnarray}
	i_{\rm max}    &=& \max [i(t_{\rm on}),i(t_p^{\rm after})] \\
	&=&    \max \bigg\{   e^{-\frac{\beta_{\rm off}}{\gamma}r(t_{\rm on})} + r(t_{\rm on}),
	\frac{\beta_{\rm off} - \beta_{\rm on}}{\beta_{\rm off}} [r(t_{\rm off})-r(t_{\rm on})]  + \frac{\gamma}{\beta_{\rm off}} \bigg[ 1 + \log \left(\frac{\beta_{\rm off}}{\gamma} \right)  \bigg]
	\bigg\}. \nonumber \\
	&&
\end{eqnarray}
%The peak $i(t_{\rm on})$ depends only on $r(t_{\rm on})$, and $i(t_p^{\rm after})$ decreases with $r(t_{\rm off})$.
%Therefore, t
The condition for the boundary $i(t_{\rm on})=i(t_p^{\rm after})$ can be rewritten as
\begin{eqnarray}
	\Delta r &=&\frac{\beta_{\rm off}}{\beta_{\rm off}-\beta_{\rm on}} \left[ e^{-\frac{\beta_{\rm off}}{\gamma}r(t_{\rm on})} + r(t_{\rm on}) \right] - \frac{\gamma}{\beta_{\rm off}-\beta_{\rm on}}\bigg[ 1 + \log \left(\frac{\beta_{\rm off}}{\gamma} \right)  \bigg]. \nonumber \\
	&&
	%\\
	%1 - e^{-\frac{\beta_{\rm off}}{\gamma}r(t_{\rm on})} - r(t_{\rm on}) 
	%    &=&
	%   \frac{\beta_{\rm off} - \beta_{\rm on}}{\beta_{\rm off}} [r(t_{\rm on})-r(t_{\rm off})] + 1 - \frac{\gamma}{\beta_{\rm off}} \bigg[ 1 + \log \left(\frac{\beta_{\rm off}}{\gamma} \right)  \bigg].
	\label{eq:boundary_i_iii}
\end{eqnarray}
This boundary is shown in Figs.~1(F) and 2(F) in the main text.

\subsection*{Boundaries between (ii) and (iii)}
Peaks at the onset and during the intervention cannot coexist. As shown in Eqs.~(\ref{eq:exist_on}) and (\ref{eq:exist_iii}),
\begin{eqnarray}
	r(t_{\rm on}) &=& \frac{\gamma}{\beta_{\rm off}} \log\left( \frac{\beta_{\rm on}}{\gamma} \right),
\end{eqnarray}
gives the boundary between regions (ii) and (iii). This condition does not depend on $\Delta r$ (Fig.~1(F) in the main text).

\subsection*{Boundaries between (iii) and (iv)}
As shown in Eqs.~(\ref{eq:exist_iii}) and (\ref{eq:exist_before}), 
\begin{eqnarray}
	r(t_{\rm on}) &=&  \frac{\gamma}{\beta_{\rm off}} \log \left( \frac{\beta_{\rm off}}{\gamma} \right),
\end{eqnarray}
gives the condition for the boundary between regions (iii) and (iv). As in the case of the boundary between (ii) and (iii), this condition does not depend on $\Delta r$ (Figs.~1(F) and 2(F) in the main text).

\section*{S4 Method: Derivation of the final size equation with the intervention}
As discussed in Eq.~(\ref{eq:s_after}), the relationship between $s(t_1)$ and $r(t_1)$ after the intervention at $t_1 > t_{\rm off}$ is
\begin{eqnarray}
	s(t_1) &=& 
	\exp \bigg[ - \frac{\beta_{\rm off}}{\gamma} r(t_1) \bigg]
	\exp\bigg\{  \frac{\beta_{\rm off}-\beta_{\rm on}}{\gamma} \bigg[ r(t_{\rm off})-r(t_{\rm on}) \bigg] \bigg\}.
\end{eqnarray}
We obtain the final size equation for $t\rightarrow \infty$ with the intervention by substituting this into the condition $s(\infty)+r(\infty) =1$ as
\begin{eqnarray}
	r(\infty) &=& 1-
	\exp \bigg[ - \frac{\beta_{\rm off}}{\gamma} r(\infty) \bigg]
	\exp\bigg\{  \frac{\beta_{\rm off}-\beta_{\rm on}}{\gamma} \bigg[ r(t_{\rm off})-r(t_{\rm on}) \bigg] \bigg\},
	\\ %\label{eq:final}
\end{eqnarray}
which is Eq.~(6) in the main text.

\section*{S5 Appendix: Features of the final size equation}
Let us study the features of Eq.~(\ref{eq:final}).
Note that we assume that $R_{0,\rm on}$ is fixed in this section.
%\textcolor{red}{$r(\infty)<r_0 (\infty)$?}

%\textcolor{red}{Minimum}
First, we show that the final size monotonically decreases with respect to $r(t_{\rm off})$ for a fixed $r(t_{\rm on})$. Regareing Eq.~(\ref{eq:final}) as an implicit function of $r(\infty)$ and $\Delta r$, we obtain
% \begin{eqnarray}
%     \frac{\partial r(\infty)}{\partial (\Delta r)} &=& \left(\frac{\beta_{\rm off}}{\gamma} \frac{\partial r(\infty)}{\partial (\Delta r)} - \frac{\beta_{\rm off}-\beta_{\rm on}}{\gamma}  \right) 
%     [1-r(\infty)],
% \end{eqnarray}
% or equivalently 
\begin{eqnarray}
	\bigg\{ 1 -\frac{\beta_{\rm off}}{\gamma} [1-r(\infty)] \bigg\} \frac{\partial r(\infty)}{\partial (\Delta r)} &=& - \frac{\beta_{\rm off}-\beta_{\rm on}}{\gamma}  
	[1-r(\infty)].
\end{eqnarray}
In the final state, the inequality $0\le 1-r(\infty)\le \gamma/\beta_{\rm off}$ must hold because of the herd immunity condition. 
Therefore, 
\begin{eqnarray}
	\frac{\partial r(\infty)}{\partial (\Delta r)}     &\le & 0,
\end{eqnarray}
holds. As $\Delta r$ monotonically increases with respect to $t_{\rm off}$ for a fixed $t_{\rm on}$, the final size is smaller if the intervention is longer.
Equivalently, $r(\infty)$ monotonically increases with respect to $r(t_{\rm off})$ for a fixed $r(t_{\rm on})$.

Note that there exists the upper bound in 
$r(t_{\rm off})$.
%Another important implication of Eq.~(\ref{eq:final}) is the existence of an upper bound in $r(t_{\rm off})$. 
Let this upper bound be $\tilde{r}(r(t_{\rm on}))$. It satisfies Eq.~(7) in the main text:
\begin{eqnarray}
	\tilde{r} &=& 1 - \exp \left[-\frac{\beta_{\rm off} - \beta_{\rm on}}{\gamma} r(t_{\rm on})\right] \exp\left[-\frac{\beta_{\rm on}}{\gamma}  \tilde{r}\right],
	\\ %\label{eq:final_intervention}
\end{eqnarray}
which is obtained for $\tilde{r}=r(\infty)$ in Eq.~(\ref{eq:final}).
This upper bound is observed in Figs.~1(F) and 2(F) in the main text.

Finally, let us discuss $r(t_{\rm on})$ and $r(t_{\rm off})$ which minimizes the final size with fixed $R_{0,\rm on}$.
The monotonicity of $r(\infty)$ with respect to $r(t_{\rm off})$ implies that the lower bound of the final size $r(\infty)$ with fixed $r(t_{\rm on})$ is obtained if  $r(t_{\rm off})$ is replaced with $ \tilde{r}(r(t_{\rm off}))$. Therefore, we  focus on the final size equation in which $r(t_{\rm off})$ is replaced with $\tilde{r}$,
\begin{eqnarray}
	r(\infty) &=& 1-
	\exp \bigg[ - \frac{\beta_{\rm off}}{\gamma} r(\infty) \bigg]
	\exp\bigg\{  \frac{\beta_{\rm off}-\beta_{\rm on}}{\gamma} \bigg[ \tilde{r}(r(t_{\rm on}))-r(t_{\rm on}) \bigg] \bigg\}.
	\label{eq:final_bound}
\end{eqnarray}
To minimize $r(\infty)$ in this equation, we should maximize $\tilde{r}(r(t_{\rm on}))-r(t_{\rm on})$ with respect to $r(t_{\rm on})$. This condition is given by 
\begin{eqnarray}
	\frac{\partial }{\partial r(t_{\rm on})} [\tilde{r} - r(t_{\rm on})]
	&=& \frac{(\beta_{\rm off}-\beta_{\rm on})(1-\tilde r)}{\gamma - \beta_{\rm on }(1-\tilde r)} -1\\
	&=& 0.
\end{eqnarray}
This can be satisfied for
\begin{eqnarray}
	\tilde r &=& 1 - \frac{\gamma}{\beta_{\rm off}} = 1 - \frac{1}{R_{0,\rm off}},\\
	r(t_{\rm on}) &=& \frac{\gamma}{\beta_{\rm off}-\beta_{\rm on}} 
	\left[\log \left(\frac{\beta_{\rm off}}{\gamma} \right) - \beta_{\rm on} \left( \frac{\beta_{\rm off}-\gamma}{\gamma \beta_{\rm off}} \right) \right].
\end{eqnarray}
By substituting these equations into Eq.~(\ref{eq:final_bound}), we obtain the final size equation:
\begin{eqnarray}
	r(\infty) &=& 1 - \frac{\gamma}{\beta_{\rm off}} \exp \bigg[ \frac{\beta_{\rm off}}{\gamma } -1 \bigg] \exp \bigg[-\frac{\beta_{\rm off}}{\gamma} r(\infty) \bigg].
\end{eqnarray}
Interestingly, this transcendental equation has the solution $r(\infty) = \tilde{r}=1 - \frac{1}{R_{0,\rm off}}$ if $R_{0,\rm on} <  R_{0,\rm off} \log \left(R_{0,\rm off} \right) / (R_{0,\rm off}-1)$ (Eq.~(8) in the main text). In other words, one can achieve herd immunity with the theoretically minimal removed fraction by intervening once if the intervention reproduction number is sufficiently small.

\section*{S6 Appendix: Numerical results for the final size}
Figures 1(B)--(E) and 2(B)--(E) in the main text present the numerical results of the final size $r(\infty)$ by varying the intervention onset time $t_{\rm on}$ with the constant basic reproduction number without the intervention $R_{0,{\rm off}}=2$ and intervention duration $\Delta t=60$ days. As suggested theoretically in S5 Appendix, the final size with the interventions (Figs.~1(B)--(E), 2(B)--(E) in the main text) is smaller than that without the intervention (Fig.~1(A) and 2(A) in the main text). 

The final size is minimized in intermediate $t_{\rm on}$ in both cases (Figs.~1(D) and 2(D) in the main  text).  This result is intuitively understood as follows. The second wave occurs if the intervention onset is early (Figs.~1(B) and 2(B) in the main text). Conversely, the removed fraction is sufficiently enough before the intervention and the second wave is not observed if the onset is late (Figs.~1(E) and 2(E) in the main text).

For sufficiently large $\Delta t$ and small enough $R_{0,\rm on}$, the final size reaches close to $1-\frac{1}{R_{0,\rm off}}=0.5$, which is the lower bound of the final size necessary to achieve herd immunity, for $r(t_{\rm on}) \approx 0.295$ (Fig.~2(D) in the main text), which corresponds to $t_{\rm on} =42.681$ days. 
It should be noted that a maximum of $\Delta r=r(t_{\rm off})-r(t_{\rm on})$ can be found for the intermediate $r(t_{\rm on})$  (Figs.~S\ref{fig:supp_final_large} (left) and S\ref{fig:supp_final_small} (left)). 
Equation (6) in the main text (Eq.~(\ref{eq:final})) suggests that a larger $\Delta r$ yields a smaller $r(\infty)$.
%The final size is minimized if $t_{\rm on}>0$ for small $\beta_{\rm on}$ as suggested by Eq.~(\ref{eq:opt_timing}) (Fig.~\ref{fig:small_beta1}(H), (I)).
Because $\tilde{r}$ is largest at intermediate $t_{\rm on}$, the final size is smallest at intermediate $t_{\rm on}$ (see S5 Appendix).

As suggested by Eq.~(\ref{eq:final}), the final size depends on $\Delta r$ but is independent of $r(t_{\rm on})$ (Figs.~S\ref{fig:supp_final_large}(A), S\ref{fig:supp_contour_large}(B), S\ref{fig:supp_final_small}(A), and S\ref{fig:supp_contour_small}(B)).

\subsubsection*{Weak intervention (large $R_{0,\rm on}=1.4$)}
\begin{figure}
	\centering
	\vspace{-4.5cm}
	\includegraphics[width=0.9\textwidth]{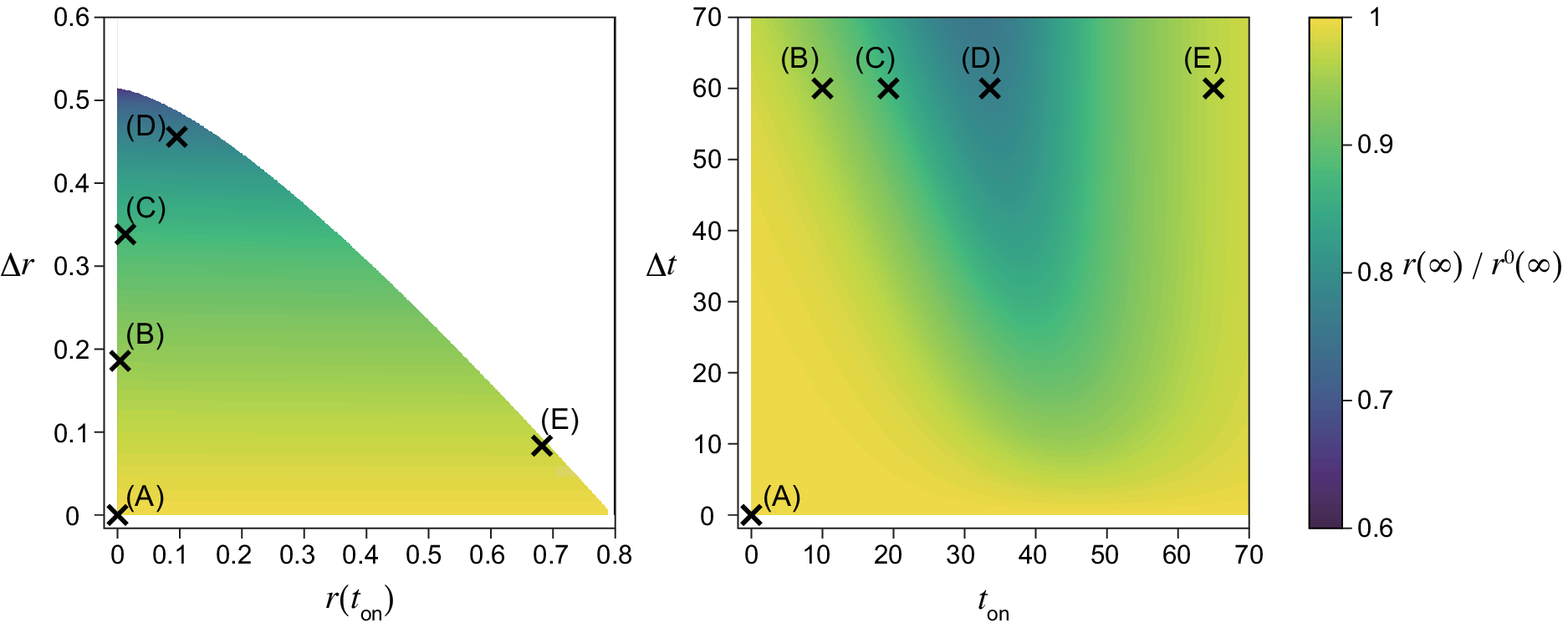}
	\caption{Final size with the intervention with large $R_{0,\rm on}=1.4$. Parameters are $\beta_{\rm off}=2/7 \ {\rm days}^{-1}$,  $\beta_{\rm on}=1.4/7 \ {\rm days}^{-1}$, $\gamma=1/7 \ {\rm days}^{-1}$, which are the same as those presented in Fig.~1 in the main text. 
		The final size of the outbreak for each case is represented by $r(\infty)$, which is normalized by that without the intervention, $r^0(\infty)$. Left and right figures are plotted in terms of $r(t_{\rm on})$ and $\Delta r$, and $t_{\rm on}$ and $\Delta t$, respectively. Symbols (A)-(E) in the figures denote the intervention timings of the time series of those corresponding to Figs.~1(A)-(E) in the main text. }
	\label{fig:supp_final_large}
\end{figure}

The final size can be minimized with $t_{\rm on} = 0$, the early implementation of the intervention, in this case (Fig.~S\ref{fig:supp_final_large}). The timing of the intervention that minimizes the final size is different from that which minimizes the maximum fraction of infected individuals. In general, the optimal timing of the intervention depends on the objective function to be minimized. The theoretical prediction in Eq.~(6) (Eq.~(\ref{eq:final})),  that the final size depends on  $\Delta r$, is numerically verified (Fig.~S\ref{fig:supp_final_large}(A)). For a constant intervention duration $\Delta t$ (Fig.~S\ref{fig:supp_final_large}(B)), the final size is minimized with intermediate $t_{\rm on}$ (Fig.~1(D) in the main text).

% \textcolor{red}{
% The final size $r(\infty)$ has to  be larger than  $ 1-\frac{\gamma}{\beta_{\rm off}}$, which is the minimum final size for the herd immunity.
% }

\subsubsection*{Strong intervention (small $R_{0,\rm on}=0.7$)}
\begin{figure}
	\centering
	\includegraphics[width=0.9\textwidth]{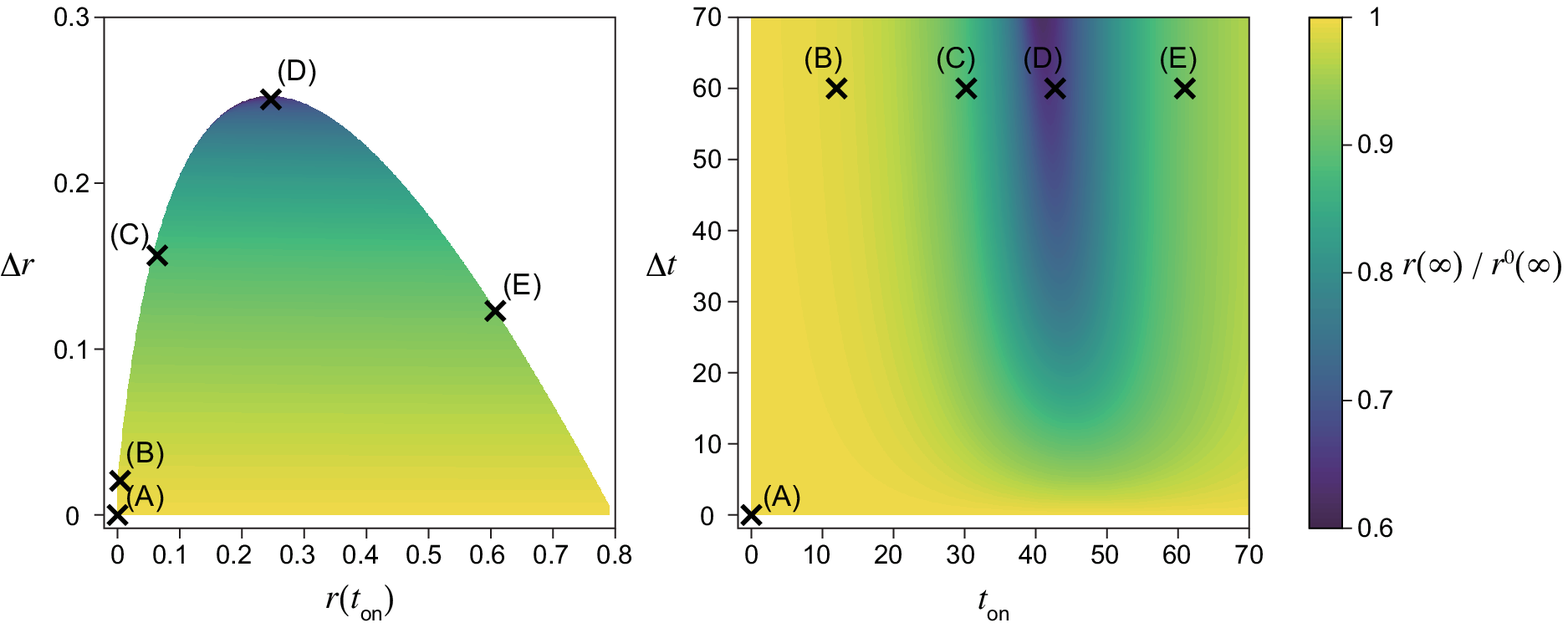}
	\caption{Final size with the intervention with a large $R_{0,\rm on}=0.7$. Parameters are $\beta_{\rm off}=2/7 \ {\rm days}^{-1}$,  $\beta_{\rm on}=0.7/7 \ {\rm days}^{-1}$, $\gamma=1/7 \ {\rm days}^{-1}$, which are the same as those in Fig.~2 in the main text. 
		The final size of the outbreak for each case is represented by $r(\infty)$, which is normalized by that without the intervention, $r^0(\infty)$. Left and right figures are plotted using $r(t_{\rm on})$ and $\Delta r$, and $t_{\rm on}$ and $\Delta t$, respectively. Symbols (A)-(E) in the figures denote the intervention timings of the time series of those corresponding to Figs.~2(A)-(E) in the main text.}
	\label{fig:supp_final_small}
\end{figure}

The simulation results (Fig.~2(B)--(E) in the main text) suggest that the final size is minimized with an intermediate starting time for small $R_{0,\rm on}$. Indeed, as shown in Fig.~S\ref{fig:supp_final_small}, the final size can be minimized at $t_{\rm on}\neq 0$. 
The minimized final size is closer to $r(\infty)=1-\frac{1}{R_{0,\rm off}}=0.5$, which is the minimum fraction of removed individuals necessary to achieve herd immunity, compared with the case of $R_{0,\rm on}=1.4$ (Fig.~2(D) in the main text). 
%In this case, the fraction of the removed individuals can be close to the largest value of $\Delta r$ for each $r(t_{\rm on})$, namely, symbols (B)-(E) reaches the edge of the reachable regions in panel (F).

\section*{S7 Figure: Contours of $i_{\rm max}$ and $r(\infty)$}
\begin{figure}
	\centering
	\includegraphics[width=0.6\textwidth]{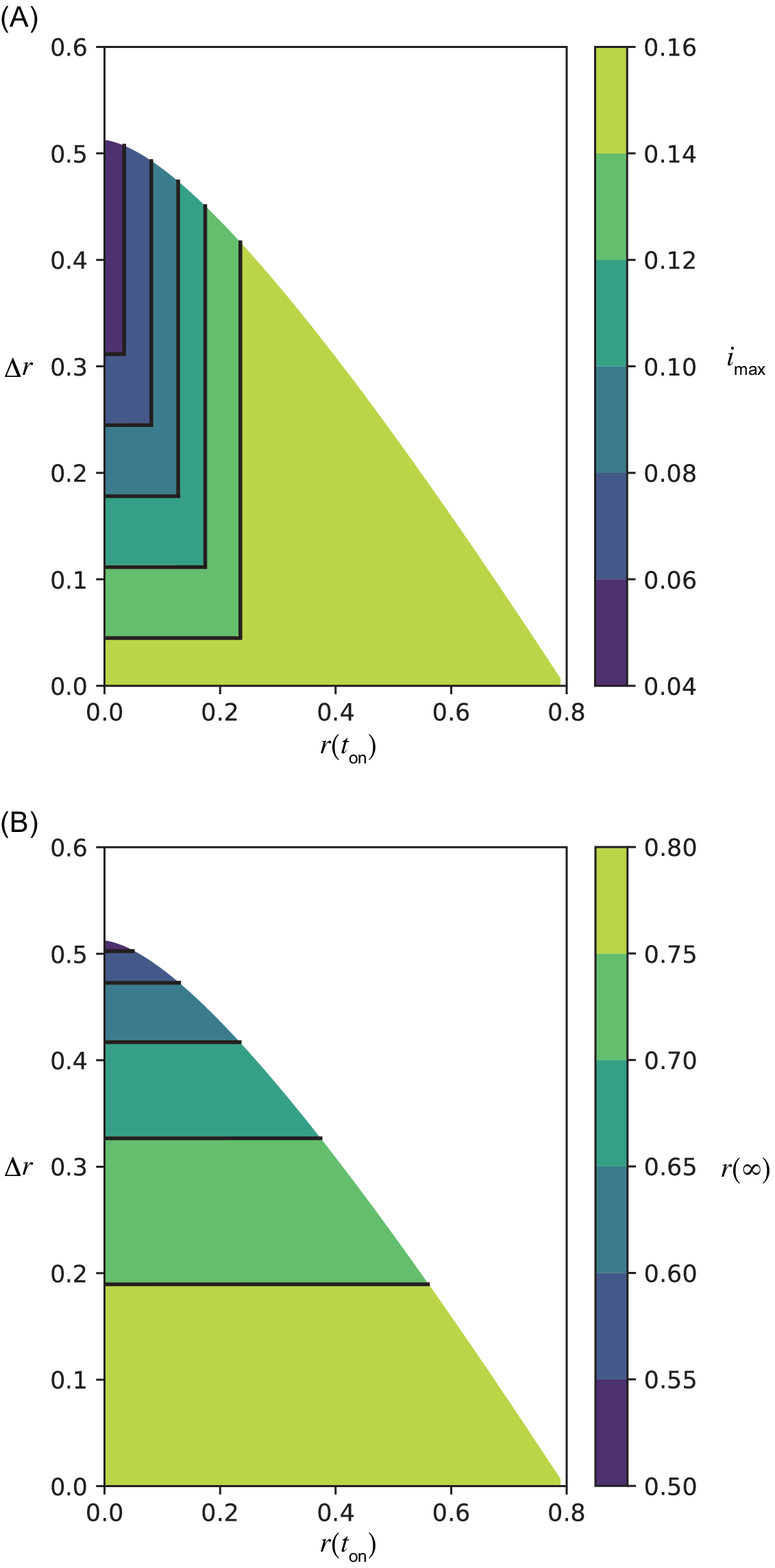}
	\caption{Contours of: (A) the maximum fraction of infected individuals, $i_{\rm max}$, and (B) the final size $r(\infty)$ for $R_{0,\rm on}=1.4$. The solid lines are obtained analytically. These figures are the same as Figs.~1(F) and (H) but with different visualizations.}
	\label{fig:supp_contour_large}
\end{figure}

\begin{figure}
	\centering
	\includegraphics[width=0.6\textwidth]{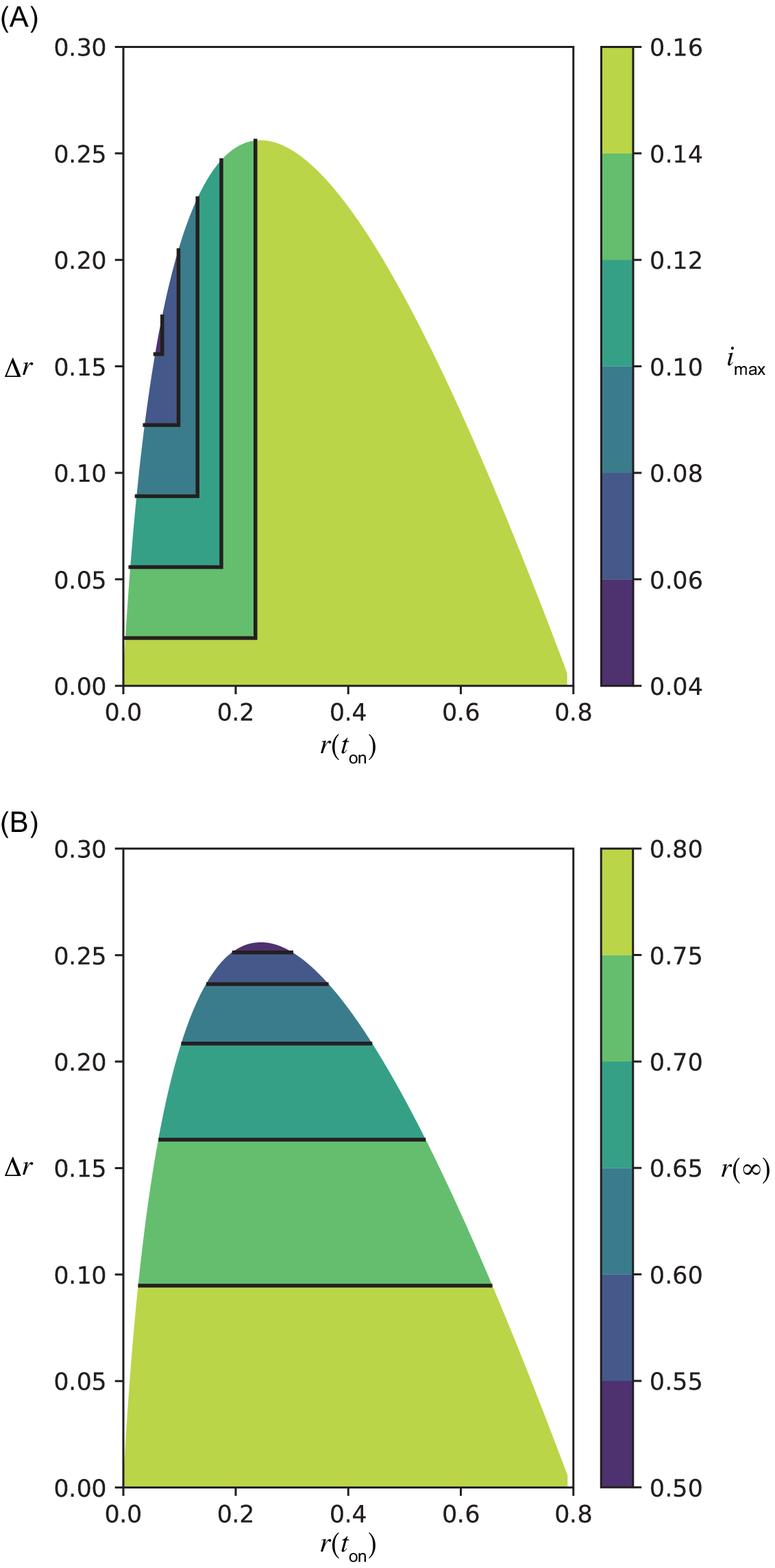}
	\caption{Contours of: (A) the maximum fraction of the infected individuals, $i_{\rm max}$, and (B) the final size, $r(\infty)$ for $R_{0,\rm on}=0.7$. The solid lines are obtained analytically. These figures are the same as Figs.~2(F) and (H) but with different visualization.}
	\label{fig:supp_contour_small}
\end{figure}

As mentioned in the main text, the maximum fraction of infected individuals $i_{\rm max}$ depends linearly on $\Delta r$ if it appears after the intervention (case (i)). If the maximum appears during or at the onset of the intervention (cases (ii) and (iii)), $i_{\rm max}$ depends on $r(t_{\rm on})$ and does not depend on $\Delta r$. Therefore, the contours in the $(r(t_{\rm on}),\Delta r)$ plane are the horizontal and vertical lines for the former and latter cases, respectively. To visualize them explicitly, Figs.~S\ref{fig:supp_contour_large}(A) and S\ref{fig:supp_contour_small}(A) plots the five contours for $i_{\rm max}$. As a theoretical prediction, they consist of horizontal and vertical lines. 

Equation (\ref{eq:final}) theoretically predicts that the final size $r(\infty)$ with the intervention depends on $\Delta r$ only, which implies that the contours are horizontal lines (Figs.~S\ref{fig:supp_contour_large}(B) and S\ref{fig:supp_contour_small}(B)).


\section*{S8 Appendix: Dependence of $\bar{i}_{\rm max}$ on $R_{0,\rm on}$}
We want to minimize the maximum fraction of infected individuals $\bar{i}_{\rm max}$ for a fixed $R_{0,\rm on}$ by varying $t_{\rm on}$ and $t_{\rm off}$. Note that $i_{\rm max}$ is decreasing function in $\Delta r$ (Eq.~(\ref{eq:peak3})) for case (i),
and $i_{\rm max}$ does not depend on $\Delta r$ in regions (ii) and (iii).
Therefore, for a fixed $r(t_{\rm on})$, $\bar{i}_{\rm max}$ is achieved either in region (ii) or (iii). Here, $\bar{i}_{\rm max}$ can be analytically or semi-analytically calculated by evaluating $i_{\rm max}$ on the boundaries between (i) and (ii) or between (i) and (iii). 
We show below that the dependence of $\bar{i}_{\rm max}$ on $R_{0, \rm on}$ changes, in which the boundary gives $\bar{i}_{\rm max}$, leading to its non-trivial dependence on $R_{0,\rm on}$.

\subsection*{Weak intervention ($R_{0,\rm on}\ge R_{0,\rm on}^\ast \approx 1.23$)}
For large $R_{0,\rm on}$, $\bar{i}_{\rm max}$ is achieved at the boundary between regions (i) and (ii), where the peaks of the infected individuals during and after the intervention are equal (Fig.~1(F) in the main text).
To discuss the parameter regions where this boundary exists, it is necessary to take into account the upper bound of removed individuals $\tilde{r}(r(t_{\rm on}))$.
%, defined by \textcolor{red}{Eq.~(7)} in the main text. 
If $r(t_{\rm off})=r(t_{\rm on})+\Delta r$ at the boundary between regions (i) and (ii) given by Eq.~(\ref{eq:equi_int_after}) is larger than $\tilde r $, this boundary is not observed at $t_{\rm on}$. In particular, it should be verified whether $\tilde{r}(0)$ is less than $\Delta r$ in Eq.~(\ref{eq:equi_int_after}) for $r(t_{\rm on})=0$. 
Let $R_{0,{\rm on}}^\ast$ be the value of $R_{0,{\rm on}}$, below which the boundary between (i) and (ii) does not exist.
%\textcolor{red}{The boundary between (i) and (ii) exists at $t_{\rm on}=0$ for large $R_{0,\rm on}$ (Fig.~1(F)).} 
%However, $\tilde{r}(0)$ decreases it disappears for smaller $R_{0,\rm on}$, and 
For $R_{0,{\rm on}}\le R_{0,{\rm on}}^\ast$,
the boundary crosses $\tilde r$ at a non-zero intervention onset time $r(t_{\rm on})$, because the boundary is decreasing in $r(t_{\rm on})$ (Eq.~(\ref{eq:equi_int_after})) and $\tilde r$ is increasing in $r(t_{\rm on})$ for small $r(t_{\rm on})$.

For $R_{0,\rm on}\ge R_{0,\rm on}^\ast$, where there is a boundary between regions (i) and (ii) at $t_{\rm on}=0$ (Fig.~1(F) in the main text), 
$\Delta r$ can be the largest at this boundary  at $t_{\rm on}=0$. Equation (\ref{eq:peak1}) suggests that a larger $\Delta r$ yields a smaller $i_{\rm max}$,
and $\bar{i}_{\rm max}$ is given by substituting $r(t_{\rm on})=0$ into Eq.~(\ref{eq:peak_int}) as follows:
\begin{eqnarray}
	\bar{i}_{\rm max}
	&=&
	1
	-\frac{1}{R_{0,\rm on}} \bigg[ 1+\log \left(R_{0,\rm on} \right) \bigg].
	\label{eq:ibar_large}
\end{eqnarray}
This equation is used as the theoretical curve for $R_{0,\rm on}\ge R_{0,\rm on}^\ast$ in Fig.~3 in the main text.
This equation implies that 
%this is an increasing function in $R_{0,\rm on}$. Namely, 
$\bar{i}_{\rm max}$ is increasing in $R_{0,\rm on}$ for $R_{0,\rm on} \ge R_{0,\rm on}^\ast$.
%This equation is used to plot $\bar{i}_{\rm max}$ in Fig.~3 in the main text for  $R_{0,\rm on}\ge R_{0,\rm on}^\ast$. 
See the next subsection for the value of $R_{0,\rm on}^\ast$.
%Analytical estimation of $R_{0,^rm on}^\ast$ is difficult. Numerically, it is computed $R_{0,\rm on}\approx 1.23$.

\subsection*{Medium-intervention ($R_{0,\rm on}^\ast \ge R_{0,\rm on} \ge R_{0,\rm on}^{(1)}\approx 1.08$)}
If $R_{0,\rm on}$ is smaller than $R_{0,\rm on}^\ast$, region (ii) does not give the global maximum of the fraction of infected individuals for $t_{\rm on}=0$. As a result, a larger $\Delta r$, which indicates a smaller $i_{\rm max}$, is realized for $t_{\rm on}\neq 0$.
The upper bound of $r(t_{\rm off})$ at the boundary between regions (i) and (ii) equals $\tilde{r}(r(t_{\rm on}))$. 
Substituting the condition for the boundary between (i) and (ii) (Eq.~(\ref{eq:equi_int_after})) in the limit $r(t_{\rm off})\rightarrow \tilde{r}$ into the condition for $\tilde{r}$ (Eq.~(\ref{eq:final_intervention})), we obtain
\begin{eqnarray}
	\tilde{r} &=& 1 - e^{-A},
\end{eqnarray}
where
\begin{eqnarray}
	A &:=&    1 + \frac{1}{R_{0,\rm off}-R_{0,\rm on}} \big[R_{0,\rm off} \log \left(R_{0,\rm on}\right) - R_{0,\rm on}  \log \left( R_{0,\rm off}\right) \big].
\end{eqnarray}
%Since $A$ consists of the parameters, this equation represents $\tilde{r}$ explicitly in terms of the parameters at the boundary.
Substituting this into Eq.~(\ref{eq:equi_int_after}) again, $r(t_{\rm on})$ at the boundary can be explicitly given as
\begin{eqnarray}
	r(t_{\rm on}) &=& \frac{ R_{0,\rm on}}{R_{0,\rm on} -R_{0,\rm off}}  \bigg[ (1- e^{-A}) - \frac{1}{R_{0,\rm on}} A \bigg].
\end{eqnarray}
Substituting these into Eq.~(\ref{eq:peak1}) or
(\ref{eq:peak2}), we obtain
\begin{eqnarray}
	\bar{i}_{\rm max} &=& e^{-A} - \frac{1}{R_{0,\rm off}-R_{0,\rm on} } \log \left( \frac{R_{0,\rm off}}{R_{0,\rm on}} \right),
	\label{eq:ibar_mid}
\end{eqnarray}
which is the theoretical curve for $R_{0,\rm on}^\ast \ge R_{0,\rm on} \ge R_{0,\rm on}^{(1)}$ in Fig.~3 in the main text.
Note that $\bar{i}_{\rm max}$ is a decreasing function in $R_{0,\rm on}$. Therefore, $\bar{i}_{\rm max}$ is minimized at $R_{0,\rm on}=R_{0,\rm on}^\ast$. 
The condition for $R_{0,\rm on}^\ast$ is  obtained by equating Eqs.~(\ref{eq:ibar_large}) and (\ref{eq:ibar_mid}).
It is difficult to calculate $R_{0,\rm on}^{\ast} $ explicitly, and it was numerically obtained as $R_{0,\rm on}^{\ast} \approx 1.23$ in the present parameter setting.

% \textcolor{blue}{***DELETE*** This can be computed by the condition $r(t_{\rm on}) = \gamma \log(\beta_{\rm on}/\gamma) /\beta_{\rm off}$. This is numerically solved.}
% \textcolor{blue}{
% Condition is $\tilde r > \frac{\gamma}{\beta_{\rm on}} A$, namely
% \begin{eqnarray}
%     \frac{\gamma}{\beta_{\rm on}} A & \le 1 - e^{-A}.
% \end{eqnarray}
% Equality gives the condition
% \begin{eqnarray}
%     A &=& W_0 \left( - \frac{\beta_{\rm on}}{\gamma}e^{-\frac{\beta_{\rm on}}{\gamma }} \right) + \frac{\beta_{\rm on}}{\gamma } 
% \end{eqnarray}
% }

\subsection*{Strong intervention ($R_{0,\rm on}\le R_{0,\rm on}^{(1)}$)}
For small $R_{0,\rm on}$, $\bar{i}_{\rm max}$ is determined differently. It is understood that a
%the case (ii) is not observed for $R_{0,\rm on}<1$. The
peak in region (ii) does not give $i_{\rm max}$ for 
%$R_{0,\rm on} \gtrsim 1$, and this is the case for
$R_{0,\rm on}\le R_{0,\rm on}^{(1)}$, where  $R_{0,\rm on}^{(1)} \approx 1.08$.

In this case, the boundary between regions (i) and (iii) satisfying $r(t_{\rm off})=\tilde{r}(r(t_{\rm on}))$ gives the condition for $\bar{i}_{\rm max}$ (Fig.~2(F) in the main text). This condition cannot be solved explicitly with respect to $\bar{i}_{\rm max}$. Therefore, we give the parametric equations for $\bar{i}_{\rm max}$ and $R_{0,\rm on}$ in terms of $r(t_{\rm on})$. 
The parametric equation for $\bar{i}_{\rm max}$ is
\begin{eqnarray}
	\bar{i}_{\rm max} &=& 1 - e^{-\frac{\beta_{\rm off}}{\gamma}r(t_{\rm on})} - r(t_{\rm on}).
	\label{eq:param_ibar}
\end{eqnarray} 
Next, we derive the parametric equation for $R_{0,\rm on}$.
To this end, let us rewrite (\ref{eq:boundary_i_iii}) as
\begin{eqnarray}
	\Delta r &=&  \frac{B(r(t_{\rm on}))}{R_{0,\rm off}-R_{0,\rm on}},
	\label{eq:equality_on_after}
\end{eqnarray}
where
\begin{eqnarray}
	B(r(t_{\rm on})) &=& R_{0,\rm off} [e^{-\frac{\beta_{\rm off}}{\gamma}r(t_{\rm on})} + r(t_{\rm on}) ]
	-  \bigg[ 1 + \log \left(R_{0,\rm off}\right)  \bigg].
\end{eqnarray}
Substituting this into Eq.~(\ref{eq:final_intervention}), we obtain a transcendental equation for $\tilde{r}$:
\begin{eqnarray}
	\tilde r &=& 1 - \exp \left[ -R_{0,\rm off} \tilde r \right] \exp \left[   B(r(t_{\rm on})) \right] ,
\end{eqnarray}
which can be solved as
\begin{eqnarray}
	\tilde r &=& 1 + \frac{1}{R_{0,\rm off}} W_{-1} \bigg(- R_{0,\rm off} \exp\left[  B(r(t_{\rm on}))-R_{0,\rm off} \right] \bigg),
\end{eqnarray}
where $W_{-1}(\cdot) $ denotes the branch of the Lambert $W$ function which gives the real value other than the principal branch.
Finally, we obtain the parametric equation for $R_{0,\rm on}$ as
\begin{eqnarray}
	R_{0,\rm on} = R_{0,\rm off} - \frac{B(r(t_{\rm on}))}{1 + \frac{1}{R_{0,\rm off}} W_{-1} \left(- R_{0,\rm off} \exp\left[  B(r(t_{\rm on}))-R_{0,\rm off} \right] \right) -r(t_{\rm on})}. \nonumber \\
	&&
	\label{eq:param_Ron}
\end{eqnarray}
Equations (\ref{eq:param_ibar}) and (\ref{eq:param_Ron}) are used to plot the theoretical curves for $R_{0,\rm on} \le R_{0,\rm on}^{(1)}$ in Fig.~3 in the main text.

% \begin{eqnarray}
%     i_{\rm max}    &=&    \max \bigg\{ 1 - e^{-\frac{\beta_{\rm off}}{\gamma}r(t_{\rm on})} - r(t_{\rm on}),
%      \frac{\beta_{\rm off} - \beta_{\rm on}}{\beta_{\rm off}} [r(t_{\rm on})-r(t_{\rm off})] + 1 - \frac{\gamma}{\beta_{\rm off}} \bigg[ 1 + \log \left(\frac{\beta_{\rm off}}{\gamma} \right)  \bigg]
%     \bigg\}
% \end{eqnarray}
%\begin{eqnarray}
%    r(t_{\rm off}) &=& r(t_{\rm on}) + \frac{\beta_{\rm off}}{\beta_{\rm off}-\beta_{\rm on}} [e^{-\frac{\beta_{\rm off}}{\gamma}r(t_{\rm on})} + r(t_{\rm on}) ]
%    - \frac{\gamma}{\beta_{\rm off}-\beta_{\rm on}} \bigg[ 1 + \log \left(\frac{\beta_{\rm off}}{\gamma} \right)  \bigg].
%    \label{eq:equality_on_after}
%\end{eqnarray}
% \begin{eqnarray}
%     r(t_{\rm off}) &=& 1 - \exp[ - \frac{\beta_{\rm off}-\beta_{\rm on}}{\gamma} r(t_{\rm on})] \exp[-\frac{\beta_{\rm on}}{\gamma} r(t_{\rm off}) ]
% \end{eqnarray}

%\textcolor{red}{
%Crossing with $\tilde r$.
%Rewrite Eq.~(\ref{eq:equality_on_after}) as
% \begin{eqnarray}
%     \beta_{\rm on} &=& \beta_{\rm off} - \frac{B(r(t_{\rm on}))}{\Delta r}.
% \end{eqnarray}
%This leads us to
%}

\section*{S9 Appendix: Optimization problems with respect to final size}

Two scenarios for the optimization problems in the hybrid nonlinear dynamical systems are presented in the main text. Here, we discuss the problems using the final size as the objective functions.

\subsection*{Minimizing final size with a constraint in intervention duration}
Targeting the final size to be minimized, we discuss two constraints based on Figs.~S\ref{fig:supp_final_large} and S\ref{fig:supp_final_small}. As in the case for the minimization problem of the maximum fraction of infected individuals, the final size is minimized at an intermediate $t_{\rm on}$ with the constraint of a constant $\Delta t$ (Figs.~1(D) and 2(D) in the main text, and Figs.~S\ref{fig:supp_final_large} (right) and S\ref{fig:supp_final_small} (right)). Note that this $t_{\rm on}$ is different from the one which minimizes $i_{\rm max}$.  This result is intuitively understood in the right panels of these figures. The contour of the final size depends on $\Delta r$. For a constant intervention duration, $\Delta r$ is largest with intermediate $r(t_{\rm on})$, which maximizes the final size with this constraint.

\subsection*{Minimizing intervention duration with a constraint on the final size}
It is possible to consider this rule to minimize the intervention duration along the contour of the final size in Figs.~S\ref{fig:supp_final_large} and S\ref{fig:supp_final_small}.  This corresponds to intervening while keeping $\Delta r$ constant. There exists an intermediate $t_{\rm on}$ that minimizes the final size in this setting as well.

%\bibliographystyle{bmc-mathphys} % Style BST file (bmc-mathphys, vancouver, spbasic).
%\bibliography{bmc_article}      % Bibliography file (usually '*.bib' )

\end{document}